\documentclass[11pt,a4paper]{article}
\usepackage{psfrag}
\usepackage{graphics}
\usepackage{graphicx}
\usepackage{float}
\usepackage{amssymb,epsfig,amsmath,euscript,array}
\usepackage{cite}
\usepackage{multicol}
\usepackage{subfig}
\usepackage{fancyhdr}
\usepackage{color}
\usepackage{scalerel,stackengine}
\usepackage{pdflscape}
\usepackage{hyperref}
\usepackage{arydshln}

\makeatletter
\@addtoreset{equation}{section}
\makeatother

\usepackage[us,24hr]{datetime}

\def\beq{\begin{equation}}
\def\eeq{\end{equation}}

\begin{document}

\setcounter{page}{1}
\thispagestyle{empty}
\begin{flushright}
MAN/HEP/2020/006 \\
\
\\ \today
\end{flushright}
\begin{center}

\hspace{150cm}

{\LARGE{\bf \boldmath The muon $g$-2 and $\Delta \alpha$ connection} \\}
\
\\
\
\
\\
{\large Alexander Keshavarzi} \\
{\small \em Department of Physics and Astronomy, The University of Manchester, Manchester M13 9PL, U.K.} \\
{\small \em Email: alexander.keshavarzi@manchester.ac.uk} \\
\
\\
{\large William J. Marciano}\\
{\small \em Department of Physics, Brookhaven National Laboratory, Upton, New York 11973, USA} \\
{\small \em Email: marciano@bnl.gov}\\
\
\\
{\large Massimo Passera} \\
{\small \em INFN Sezione di Padova, Via Francesco Marzolo 8, 35131 Padova, Italy} \\
{\small \em Email: passera@pd.infn.it}\\
\
\\
{\large Alberto Sirlin}\\
{\small \em Department of Physics, New York University, 726 Broadway, New York, New York 10003, USA} \\
{\small \em Email: alberto.sirlin@nyu.edu}
\
\\
\
\\
\
\\

{\normalsize \bf Abstract}
\end{center}

The discrepancy between the Standard Model theory and experimental measurement of the muon magnetic moment anomaly, $a_{\mu}=\left(g_{\mu}-2\right)/2$, is connected to precision electroweak (EW) predictions via their common dependence on hadronic vacuum polarization effects. The same data for the total $e^+e^- \rightarrow \text{hadrons}$ cross section, $\sigma_{\rm had}(s)$, are used as input into dispersion relations to estimate the hadronic vacuum polarization contributions, $a_{\mu}^{\rm had,\,VP}$, as well as the five-flavor hadronic contribution to the running QED coupling at the $Z$-pole, $\Delta\alpha_{\rm had}^{(5)}(M_{Z}^2)$, which enters natural relations and global EW fits. The EW fit prediction of $\Delta\alpha_{\rm had}^{(5)}(M_{Z}^2) = 0.02722(41)$ agrees well with $\Delta\alpha_{\rm had}^{(5)}(M_{Z}^2) = 0.02761(11)$ obtained from the dispersion relation approach, but exhibits a smaller central value suggestive of a larger discrepancy $\Delta a_{\mu}=a_{\mu}^{\rm exp} - a_{\mu}^{\rm SM}$ than currently expected. Postulating that the $\Delta a_{\mu}$ difference may be due to unforeseen missing $\sigma_{\rm had}(s)$ contributions, implications for $M_W$, $\sin^2 \! \theta^{\rm lep}_{\rm eff}$ and $M_H$ obtained from global EW fits are investigated. Shifts in $\sigma_{\rm had}(s)$ needed to bridge $\Delta a_{\mu}$ are found to be excluded above $\sqrt{s} \gtrsim 0.7$~GeV at the 95\%CL. Moreover, prospects for $\Delta a_{\mu}$ originating below that energy are deemed improbable given the required increases in the hadronic cross section. Such hypothetical changes to the hadronic data are also found to affect other related observables, such as the electron anomaly, $a_e^{\rm SM}$, the rescaled ratio $R_{e/\mu} = (m_\mu/m_e)^2 (a_{e}^{\rm had,\,LO\,VP}/a_{\mu}^{\rm had,\,LO\,VP})$ and the running of the weak mixing angle at low energies, although the consequences of these are currently less constraining.

\newpage

\tableofcontents

\section{Introduction}

The muon magnetic moment anomaly, $a_{\mu} = (g-2)/2$, exhibits a long-standing discrepancy between the Standard Model (SM) prediction~\cite{Aoyama:2020ynm} and the experimentally measured value~\cite{Bennett:2002jb,Bennett:2004pv,Bennett:2006fi,PDG2018}. If confirmed with high significance, it would be indirect evidence for new physics beyond SM expectations~\cite{Czarnecki:2001pv}. Of interest in this paper is the validity and impact of this discrepancy on other well-determined physics observables, in particular those resulting from the global fit of the SM electroweak (EW) sector. These two fields of precision physics are connected via the total $e^+e^- \rightarrow \text{hadrons}$ cross section, $\sigma_{\rm had} (s)$. This cross section is currently an indispensable tool to determine both the hadronic vacuum polarization contributions to the muon anomaly, $a_{\mu}^{\rm had,\,VP}$, and the hadronic contributions to the running QED coupling at the scale of the $Z$ boson mass, $\Delta\alpha_{\rm had}^{(5)}(M_{Z}^2)$. Detailed implications of this connection were first explored in~\cite{Passera:2008jk,Passera:2008hj,Passera:2010ev} and more recently used to compare results from dispersion relations and lattice gauge theory calculations~\cite{Crivellin:2020zul}. However, since~\cite{Passera:2008jk,Passera:2008hj,Passera:2010ev}, both the SM prediction for $a_{\mu}$ and the EW fits have improved substantially, motivating the updated analysis presented here.

The present most precise measurements of $a_{\mu^+}$ and $a_{\mu^-}$ were carried out by the E821 experiment at the Brookhaven National Laboratory (BNL)~\cite{Bennett:2002jb,Bennett:2004pv,Bennett:2006fi}, resulting in a world average (assuming CPT invariance) of~\cite{PDG2018} 
\beq \label{amuExp}
a_{\mu}^{\rm exp} = (11\ 659 \ 209.1 \pm 6.3) \times 10^{-10}\,.
\eeq
Currently, a new measurement of $a_{\mu}^{\rm exp}$ is underway at the Fermilab Muon $g$-2 (E989) experiment, which is mid-way through its overall data-taking period~\cite{Grange:2015fou,Keshavarzi:2019bjn}. Its results are expected to reach a sensitivity four-times better than the E821 result. An alternative low-energy approach at J-PARC is expected to reach a precision similar to the existing BNL measurement~\cite{Abe:2019thb}. Both experiments also intend to apply their methodology to search for (and set limits on) a muon electric dipole moment (EDM).

The quoted value for the SM $a_{\mu}$ prediction by the Muon $g$-2 Theory Initiative is~\cite{Aoyama:2020ynm}
\beq \label{amuSMfinal}
a_{\mu}^{\rm SM} = (11\ 659 \ 181.0 \pm 4.3) \times 10^{-10} \, ,
\eeq
the result of which is based on~\cite{Aoyama:2012wk,Aoyama:2019ryr,Czarnecki:2002nt,Gnendiger:2013pva,Davier:2017zfy,Keshavarzi:2018mgv,Colangelo:2018mtw,Hoferichter:2019gzf,Davier:2019can,Keshavarzi:2019abf,Kurz:2014wya,Melnikov:2003xd,Masjuan:2017tvw,Colangelo:2017fiz,Hoferichter:2018kwz,Gerardin:2019vio,Bijnens:2019ghy,Colangelo:2019uex,Blum:2019ugy,Colangelo:2014qya}. As reported in~\cite{Aoyama:2020ynm}, this results in the difference
\beq
\Delta a_{\mu} = a_{\mu}^{\rm exp} - a_{\mu}^{\rm SM} = (27.9 \pm 7.6)\times 10^{-10}\, , 
\eeq
corresponding to a discrepancy of $3.7\sigma$.
The SM prediction for the muon anomaly is determined from the sum
\beq \label{alSMeq}
a_{\mu}^{\rm SM} = a_{\mu}^{\rm QED} + a_{\mu}^{\rm EW} + a_{\mu}^{\rm had,\,LbL} + a_{\mu}^{\rm had,\,VP} \, ,
\eeq
where $a_{\mu}^{\rm QED}$~\cite{Aoyama:2012wk,Aoyama:2017uqe,Aoyama:2019ryr} and $a_{\mu}^{\rm EW}$~\cite{Czarnecki:1995wq,Czarnecki:1995sz,Czarnecki:2002nt,Gnendiger:2013pva} are the QED and EW contributions, respectively. Both of these quantities have been calculated to high order in perturbation theory and have been cross-checked analytically and numerically~\cite{Kataev:2012kn,Baikov:2013ula,Kurz:2013exa, Kurz:2015bia,Kurz:2016bau,Laporta:2017okg,Volkov:2017xaq,Peris:1995bb,Czarnecki:2017rlm,Ishikawa:2018rlv}. As such, they are widely considered non-controversial. The hadronic contributions, on the other hand, cannot be reliably calculated perturbatively and rely on experimental data as input to dispersion relations, theory models and, more recently, lattice QCD (LQCD). In recent years, major progress has been made in determining the hadronic light-by-light (LbL) contribution, $a_{\mu}^{\rm had,\,LbL}$, from dispersive approaches and from LQCD. The latest data-driven and dispersive hadronic LbL results~\cite{Colangelo:2014dfa, Colangelo:2014pva, Colangelo:2017qdm, Colangelo:2015ama, Colangelo:2017fiz, Masjuan:2017tvw, Hoferichter:2018kwz, Hoferichter:2018dmo, Gerardin:2019vio, Bijnens:2019ghy, Colangelo:2019uex, Pauk:2014rta, Danilkin:2016hnh, Jegerlehner:2017gek, Knecht:2018sci, Eichmann:2019bqf, Roig:2019reh, Colangelo:2014qya} and first complete LQCD evaluation~\cite{Blum:2019ugy}, confirm the previously accepted model-based `Glasgow consensus' result~\cite{Prades:2009tw}, thereby eliminating the hadronic LbL sector as the source of the muon $g$-2 discrepancy. This leaves the hadronic vacuum polarization (VP) contributions, $a_{\mu}^{\rm had,\,VP}$, as the remaining SM candidate to explain $\Delta a_\mu$. For a thorough report on the various $a_{\mu}^{\rm SM}$ contributions and their quoted values by the Muon $g$-2 Theory Initiative, see~\cite{Aoyama:2020ynm}.

The $a_{\mu}^{\rm had,\,VP}$ contributions are most precisely determined using a compilation of all available $e^+e^- \rightarrow \text{hadrons}$ cross section data, $\sigma_{\rm had}(s) \equiv \sigma^0\left(e^+e^-\rightarrow \gamma^* \rightarrow \text{hadrons} + \left(\gamma\right)\right) $, where the cross section is bare, i.e.\ excluding all vacuum polarization effects (as indicated by the superscript `0') and other higher order effects except for inclusive final state bremsstrahlung. At leading order (LO), these data are input into the dispersion relation:
\beq \label{eq:amu}
a_{\mu}^{\rm had,\,LO\,VP} =\frac{1}{4\pi^3}\int^{\infty}_{s_{th}} {\rm d}s \, K(s)\, \sigma_{\rm had}(s) \,,
\eeq
where $s_{th} = m_{\pi^0}^2$ and $K(s)$ is a well-known kernel function, given by~\cite{Brodsky:1967sr,Lautrup:1969fr}
\beq \label{eq:kernel}
K(s) = \int^{1}_{0} {\rm d}x \, \frac{x^2(1-x)}{x^2+(1-x)(s/m_\mu^2)} \,.
\eeq
Similar dispersion integrals and kernel functions allow for the determination of the next-to-leading order (NLO) contribution, $a_{\mu}^{\rm had,\,NLO\,VP} $, in exactly the same approach as in the LO case~\cite{Krause:1996rf,Hagiwara:2003da}. The KNT19 analysis~\cite{Keshavarzi:2019abf} of the hadronic VP contributions (see e.g.~\cite{Davier:2003pw,Hagiwara:2003da,Davier:2007ua,Hagiwara:2006jt,Jegerlehner:2006ju,Davier:2010nc,Jegerlehner:2011ti,Hagiwara:2011af,Jegerlehner:2017lbd,Jegerlehner:2017gek,Davier:2017zfy,Keshavarzi:2018mgv,Davier:2019can} for similar analyses) found these to be
\begin{align}\label{eq:KNT19HVP}
a_{\mu}^{\rm had,\,LO\,VP}[{\rm KNT19}] & = (692.78 \pm 2.42) \times 10^{-10} \, , \nonumber 
\\
\
a_{\mu}^{\rm had,\, NLO \,VP}[{\rm KNT19}] & = (-9.83 \pm 0.04) \times 10^{-10} \, .
\end{align}
These results are essential inputs to the value of $a_{\mu}^{\rm SM}$ presented by the Muon $g$-2 Theory Initiative~\cite{Aoyama:2020ynm} and given in equation~\eqref{amuSMfinal}. Coupled with the estimate of the NNLO contribution, $a_{\mu}^{\rm had,\, NNLO\, VP} = 1.24(1) \times 10^{-10}$~\cite{Kurz:2014wya}, the total hadronic VP contribution to $a_\mu$ is estimated to be \allowbreak $a_{\mu}^{\rm had,\, VP}[{\rm KNT19}] = 684.19 (2.38) \times 10^{-10}$ These updates yielded $a_{\mu}^{\rm SM}[{\rm KNT19}] = (11\ 659 \ 181.1 \pm 3.8) \times 10^{-10}$ and $\Delta a_{\mu}[{\rm KNT19}] = (28.0 \pm 7.4)\times 10^{-10}$~\cite{Keshavarzi:2019abf}, which are entirely consistent with the corresponding values presented in~\cite{Aoyama:2020ynm} and form the basis of the results presented in this paper.~\footnote{From this point onward, all references to $a_{\mu}^{\rm SM}$ and $\Delta a_{\mu}$ correspond to $a_{\mu}^{\rm SM}[{\rm KNT19}]$ and $\Delta a_{\mu}[{\rm KNT19}]$, respectively.}

The same hadronic cross section data are also used to determine the five-flavor hadronic contributions to the running QED coupling $\alpha(q^2)$ evaluated at the mass of the $Z$ boson, $\Delta\alpha_{\rm had}^{(5)}(M_{Z}^2)$.\footnote{The full running of $\alpha(q^2)$ from leptons (lep), the five-flavor hadronic contributions and the top quark is defined as $\alpha(q^2)=\alpha/\left(1-\Delta\alpha(q^2)\right)$, where $\Delta\alpha(q^2)=\Delta\alpha_{\rm lep}(q^2) +\Delta\alpha_{\rm had}^{(5)}(q^2) + \Delta\alpha_{\rm top}(q^2)$.
The $W$-loop vacuum polarization effects, which are generally excluded in the on-shell definition, are not explicitly included here. Throughout this paper, $\alpha(q^2)$ and $\Delta\alpha(q^2)$ are considered for time-like $q^2>0$. Translating to space-like $q^2<0$, or lattice Euclidean space, which will not be considered here, is needed for $t$-channel QED processes.} They are determined by the dispersion relation
\beq \label{eq:delAlpha}
\Delta\alpha_{\rm had}^{(5)}(M_{Z}^2) \, = \, \frac{M_{Z}^2}{4\alpha\pi^2} \,\,{\rm P} \!
\int^{\infty}_{s_{th}} {\rm d}s\frac{\sigma_{\rm had} (s)}{M_{Z}^2-s}\,,
\eeq
where $\alpha$ is the fine-structure constant and P indicates the principal value of the integral. 
The KNT19 analysis~\cite{Keshavarzi:2019abf} (see also~\cite{Jegerlehner:2017lbd,Davier:2019can}) found
\beq \label{delalphahad_KNT19}
\Delta\alpha_{\rm had}^{(5)}(M_Z^2)[{\rm KNT19}] = (276.09 \pm 1.12) \times 10^{-4} \ .
\eeq
Here, and throughout this paper, $\Delta\alpha_{\rm had}^{(5)}(M_{Z}^2)$ is evaluated at time-like $q^2 = M_Z^2$. This quantity is an important ingredient for global EW fits, where precision measurements of EW observables are compared to accurate predictions of various parameters of the EW sector of the SM~\cite{Flacher:2008zq, Baak:2011ze, Baak:2012kk, Baak:2014ora, Haller:2018nnx, Alcaraz:2006mx, Alcaraz:2007ri, Group:2008aa, Group:2009ae, ALEPH:2010aa, deBlas:2016ojx, deBlas:2019okz}. Prior to its discovery, these fits predicted and set bounds on the mass of the Higgs boson, $M_H$. With the Higgs now firmly established in the SM~\cite{Aad:2012tfa, Chatrchyan:2012xdj}, the EW fits have become more constrained and, in some cases, the prediction of a parameter is now more precise than its measurement~\cite{Haller:2018nnx}. This is true for the mass of the $W$ boson, $M_W$, and for the effective EW mixing angle, $\sin^2 \! \theta^{\rm lep}_{\rm eff}$. All three parameters, $M_W$, $\sin^2 \! \theta^{\rm lep}_{\rm eff}$ and $M_H$ depend on $\Delta\alpha_{\rm had}^{(5)}(M_Z^2)$ and, therefore, their EW predictions rely upon the hadronic cross section data utilized in equation~\eqref{eq:delAlpha}.

Therein lies the connection between the muon $g$-2 and the EW sector of the SM. Assuming that the muon $g$-2 discrepancy originates from the hadronic VP contributions due to some hypothetical missed contribution in $\sigma_{\rm had}(s)$, this contribution would also be missing from the input to $\Delta\alpha_{\rm had}^{(5)}(M_{Z}^2)$. As $\Delta\alpha_{\rm had}^{(5)}(M_{Z}^2)$ is important for EW precision fits, any additional contribution must also alter the predicted values of the EW observables. This work investigates the possibility of such a claim and, specifically, focuses on the effect on the predicted values of $M_W$, $\sin^2 \! \theta^{\rm lep}_{\rm eff}$ and $M_H$ when the hadronic cross section data are adjusted to account for the muon $g$-2 discrepancy. As part of this analysis, the implications for other observables connected via the hadronic vacuum polarization sector are also of interest. The electron $g$-2 ($a_e$), the ratio \allowbreak $R_{e/\mu} = (m_\mu/m_e)^2(a_{e}^{\rm had,\,LO\,VP}/a_{\mu}^{\rm had,\,LO\,VP})$ and the running of the weak mixing angle at low energies all contain hadronic contributions which depend on $\sigma_{\rm had}(s)$.

Recently, the BMW collaboration presented the first LQCD determination of $a_{\mu}^{\rm had,\,LO\,VP}$ with sub-percent (0.6\%) precision~\cite{Borsanyi:2020mff}. This impressive result, $a_{\mu}^{\rm had,\,LO\,VP}[{\rm BMW}] = 712.4 (4.5) \times 10^{-10}$, leads to a value for $a_{\mu}^{\rm SM}$ that is in good agreement with equation~\eqref{amuExp}, therefore eliminating the muon $g$-2 discrepancy and indicating a no-new-physics scenario.\footnote{Following the publication of this paper, the BMW collaboration presented the updated value $a_{\mu}^{\rm had,\,LO\,VP}[{\rm BMW}] = 708.7(5.3) \times 10^{-10}$.} However, this result is in tension with the dispersive evaluations, being $3.8\sigma$ higher than $a_{\mu}^{\rm had,\,LO\,VP}[{\rm KNT19}]$. The implications of the BMW evaluation for the EW sector of the SM were recently investigated in~\cite{Crivellin:2020zul} following arguments similar to those made in~\cite{Passera:2008jk,Passera:2008hj,Passera:2010ev} and described above. The conclusions from~\cite{Crivellin:2020zul}, obtained using energy-independent increases of $\sigma_{\rm had}(s)$ in different energy ranges, were that accounting for the BMW determination creates tensions within the global EW fit (that may be due to new physics effects in the global fit). However, as the source of unforeseen missing contributions to the hadronic cross section may be highly energy-dependent, fully incorporating the different energy-dependent weighting of the dispersion integrals for $a_{\mu}^{\rm had,\,LO\,VP}$ and $\Delta\alpha_{\rm had}^{(5)}(M_{Z}^2)$ is essential when investigating the corresponding effect on the EW sector of the SM. Therefore, as in~\cite{Passera:2008jk, Passera:2008hj, Passera:2010ev}, shifts to $\sigma_{\rm had}(s)$ in this work are investigated in a fully energy-dependent approach, which will prove imperative for understanding and contrasting the relationship between the dispersive approach and the BMW LQCD result.\footnote{For the major part of this work, discussions concerning shifting $\sigma_{\rm had}(s)$ to account for $\Delta a_\mu$ will be made in comparison with $a_{\mu}^{\rm exp}$. An extensive comparison with the lattice results of the BMW collaboration~\cite{Borsanyi:2020mff} is beyond the scope of this paper. For an up-to-date discussion, see~\cite{Aoyama:2020ynm} and~\cite{Lehner:2020crt}.}

The structure of this paper is as follows. Section~\ref{sec:ChangesSincePreviousAnalysis} describes the process by which the hadronic data are adjusted to account for $\Delta a_{\mu}$ and details the major updates and improvements to the hadronic cross section data, the measurements of the EW observables and the EW fit used in this work compared to the previous analysis~\cite{Passera:2008jk}. Section~\ref{sec:Results} focuses on the corresponding results for $M_W$, $\sin^2 \! \theta^{\rm lep}_{\rm eff}$ and $M_H$ when shifting $\sigma_{\rm had}(s)$ to account for $\Delta a_\mu$, and discusses the plausibility of this hypothesis with respect to the required changes in the measured cross section. Section~\ref{sec:reverseArgument} details a reversal of the original argument, where the EW fit is used to provide a prediction for $\Delta\alpha_{\rm had}^{(5)}(M_{Z}^2)$ and describes the impact of this on the muon $g$-2 discrepancy. The impact of these changes on other connected observables is explored for the electron $g$-2 and the ratio of the electron/muon vacuum polarization contributions ($R_{e/\mu}$) in Section~\ref{sec:a_e}, and for the weak mixing angle at low energies in Section~\ref{sec:weakangle}. Conclusions are summarized in Section~\ref{Conclusions}.

\section{Changes since the previous analysis} \label{sec:ChangesSincePreviousAnalysis}

\subsection{Methodology summary and updates}\label{sec:dAlphaShiftMethod}

Following the methodology in~\cite{Passera:2008jk}, the definitions 
\begin{align}
\label{eq:a}
a = a_{\mu}^{\rm had,\,VP}\bigg|_{s < M_Z^2} & = \int^{M_Z^2}_{s_{th}}{\rm d}s\,f(s)\,\sigma_{\rm had} (s) \,,
\\
\
\label{eq:b}
b = \Delta\alpha_{\rm had}^{(5)}(M_{Z}^2)\bigg|_{s < M_Z^2} & = \int^{M_Z^2}_{s_{th}}{\rm d}s\,g(s)\,\sigma_{\rm had} (s) \,,
\end{align}
are adopted from equations~\eqref{eq:amu} and~\eqref{eq:delAlpha}, respectively, where $g(s) = [M_{Z}^2/(M_{Z}^2-s)]/(4\alpha\pi^2)$. As an improvement to~\cite{Passera:2008jk}, $f(s)$ now includes the dominant NLO contributions to $a_{\mu}^{\rm had,\,VP}$. These NLO contributions (that enter at $\mathcal{O}(\alpha^3)$) can be conveniently divided into three classes as defined in~\cite{Krause:1996rf}: 2(a) contains those higher order corrections from an additional virtual photon or a muon loop, 2(b) contains diagrams with an additional electron or $\tau$ loop, and 2(c) contains a second hadronic insertion in addition to the leading order hadronic contribution. Noting that the 2(c) contributions require a double integral over $\sigma_{\rm had}(s)$ and $\sigma_{\rm had}(s')$ with a kernel function $K(s,s')$ and that they only account for $\sim 0.05\%$ of $a_{\mu}^{\rm had, \,VP}$, here the 2(c) contributions are excluded from $a_{\mu}^{\rm had, \, NLO \,VP}$. This approximation is supported by the knowledge that they will provide negligible changes to the variation in $\Delta\alpha_{\rm had}^{(5)}(M_{Z}^2)$.\footnote{As the NNLO contributions to $a_{\mu}^{\rm had, \,VP}$ are of the same order as the 2(c) NLO contributions, they are also safely excluded from this analysis.} Therefore, for the purpose of this work,
\begin{align}
a_{\mu}^{\rm had,\,VP} & \approx a_{\mu}^{\rm had, \, LO \,VP} + a_{\mu}^{{\rm had, \, NLO \,VP}, \, 2(a)} + a_{\mu}^{{\rm had, \, NLO \,VP}, \, 2(b)} \nonumber
\\
\ 
& = \int^{\infty}_{s_{th}}{\rm d}s\, \left[\frac{1}{4\pi^3}K(s) + \frac{\alpha}{4\pi^4}K^{2(a)}(s) + \frac{\alpha}{4\pi^4}K^{2(b)}(s) \right]\sigma_{\rm had} (s) \, ,
\end{align}
such that 
\beq
f(s) = \frac{1}{4\pi^3}\left[K(s) + \frac{\alpha}{\pi}\left(K^{2(a)}(s) + K^{2(b)}(s) \right) \right] \, .
\eeq

The behaviors of the kernels $f(s)$ and $g(s)$ are very different. For energies much lower than $M_Z$, $g(s)$ is roughly constant. On the contrary, $f(s)$ decreases monotonically for increasing $s$ and, for large $s$, it behaves as $m_{\mu}^2/(3s)$ to a good approximation. Therefore, the weight function in the $a_{\mu}^{\rm had,\,VP}$ integral gives a stronger weight to low-energy data than the weight function for $\Delta\alpha_{\rm had}^{(5)}(M_{Z}^2)$. If the assumption is made that the muon $g$-2 discrepancy is due to a missed contribution in $\sigma_{\rm had} (s) $, then, because of the difference in the kernel functions $f(s)$ and $g(s)$, the magnitudes of these missed contributions to $a_{\mu}^{\rm had,\,VP}$ and $\Delta\alpha_{\rm had}^{(5)}(M_{Z}^2)$ will be different and dependent on the energy region.

As in~\cite{Passera:2008jk}, energy-dependent increases to the cross section are applied in two ways: 
\begin{enumerate}
\item Increases of a multiplicative, positive constant $\epsilon$ in the form 
\beq
\Delta\sigma(s) = \epsilon\sigma(s)\, ,
\eeq
over an energy window $(\sqrt{s_0} - \delta/2) \leq \sqrt{s} \leq (\sqrt{s_0} + \delta/2)$. Here, $\sqrt{s_0}$ defines the center of the energy region of width $\delta$, and $(m_{\pi^0} + \delta/2) < \sqrt{s_0} < (M_Z - \delta/2)$. Assuming an increase $\Delta a(\sqrt{s_0},\delta,\epsilon)$ to $a_{\mu}^{\rm had,\,VP}$ that is equal to the muon $g$-2 discrepancy $\Delta a_\mu$, the parameter $\epsilon$ is given by
\beq
\epsilon = \frac{\Delta a_\mu}{\int^{\sqrt{s_0} + \delta/2}_{\sqrt{s_0} - \delta/2}2E\,f(E^2)\,\sigma(E^2)\,{\rm d}E}\, ,
\eeq
where $E = \sqrt{s}$. This, in turn, induces a corresponding increase $\Delta b(\sqrt{s_0},\delta)$ to $\Delta\alpha_{\rm had}^{(5)}(M_{Z}^2)$ of
\beq \label{eq:DeltabBinned}
\Delta b(\sqrt{s_0},\delta) = \Delta a_\mu\frac{\int^{\sqrt{s_0} + \delta/2}_{\sqrt{s_0} - \delta/2}2E\,g(E^2)\,\sigma(E^2)\,{\rm d}E}{\int^{\sqrt{s_0} + \delta/2}_{\sqrt{s_0} - \delta/2}2E\,f(E^2)\,\sigma(E^2)\,{\rm d}E}\, .
\eeq
\item Point-like increases, defined as
\beq
\Delta\sigma(s) = \epsilon'\delta(s-s_0)\, ,
\eeq
where $m_{\pi^0} < \sqrt{s_0} < M_Z $. In this scenario, an increase $\Delta a(\sqrt{s_0},\epsilon') = \Delta a_{\mu} = \epsilon'f(s_0)$ results in an increase to $\Delta\alpha_{\rm had}^{(5)}(M_{Z}^2)$ of
\beq \label{eq:DeltabPointlike}
\Delta b(\sqrt{s_0}) = \Delta a_\mu\frac{g(s_0)}{f(s_0)}\, .
\eeq
\end{enumerate}
In both cases, the shifted value of $\Delta\alpha_{\rm had}^{(5)}(M_{Z}^2)$ is then calculated as
\beq \label{eq:DeltaAlphaNew}
\left( \Delta\alpha_{\rm had}^{(5)}(M_{Z}^2) \right)' \, = \, \Delta\alpha_{\rm had}^{(5)}(M_{Z}^2) \, + \, \Delta b\, .
\eeq
$\Delta b$ is taken from either equation~\eqref{eq:DeltabBinned}, for increases in energy bins, or equation~\eqref{eq:DeltabPointlike} for point-like increases. The uncertainty of equation~\eqref{eq:DeltaAlphaNew} is defined from the point-like scenario. It must account for the correlation between the identical hadronic cross section data used to determine $\Delta\alpha_{\rm had}^{(5)}(M_{Z}^2)$ and $a_{\mu}^{\rm had,\,VP}$, which contributes to $\Delta a_\mu$ in $\Delta b$. Therefore, defining $\Delta a_\mu = a_{\mu}^{\rm exp} - a_{\mu}^{\rm had,\,VP} - a_{\mu}^{\rm non-HVP}$, where $a_{\mu}^{\rm non-HVP} $ are all the contributions to $a_{\mu}^{\rm SM}$ other than those from the hadronic vacuum polarization sector, the uncertainty of equation~\eqref{eq:DeltaAlphaNew} is given by
\beq
\delta \! \left( \Delta\alpha_{\rm had}^{(5)}(M_{Z}^2) \right)' \! = \left\{ \! \left[\delta\Delta\alpha_{\rm had}^{(5)}(M_{Z}^2) - \frac{g(s_0)}{f(s_0)}\delta a_{\mu}^{\rm had,\,VP} \right]^2 \! + \left(\frac{g(s_0)}{f(s_0)} \right)^{\!2} \!\! \left[ \left(\delta a_{\mu}^{\rm exp}\right)^2 + \left(\delta a_{\mu}^{\rm non-HVP}\right)^2 \right] \! \right\}^\frac{1}{2} \!\! ,
\eeq
where the first term accounts for the correlation between $\Delta\alpha_{\rm had}^{(5)}(M_{Z}^2)$ and $a_{\mu}^{\rm had,\,VP}$.

\subsection{Updates to $\sigma_{\rm had} (s)$}

In this work, the total hadronic cross section data as determined in the KNT19 analysis~\cite{Keshavarzi:2019abf} are used.\footnote{Historically (and explored in~\cite{Passera:2008jk}), $a_{\mu}^{\rm had,\,VP}$ has also been computed incorporating hadronic $\tau$-decay data, which were related to the $e^+e^-\rightarrow$ hadrons cross section via isospin symmetry~\cite{ADH98, Davier:2009ag, Davier:2003pw, Hagiwara:2003da, Davier:2007ua, Hagiwara:2006jt, Jegerlehner:2006ju, Davier:2010nc, Jegerlehner:2011ti, Jegerlehner:2017lbd, Jegerlehner:2017gek,IVC1}. However, as the isospin-breaking corrections that must be applied are currently disputed (see~\cite{Aoyama:2020ynm} for further details), and also noting that the vast catalogue of available $e^+e^-$ data has now far surpassed their precision without the need for additional theory-based corrections, they will not be featured in this work.} These data are significantly improved in comparison to the data that were available in~\cite{Passera:2008jk}. In particular, the volume of available data since that time has increased dramatically, incorporating many new measurements of previously included hadronic modes and a large number of previously unmeasured hadronic modes.\footnote{Details and in-depth discussions of the updates to the data compilation of $\sigma_{\rm had} (s)$ used here can be found in~\cite{Hagiwara:2006jt,Hagiwara:2011af,Keshavarzi:2018mgv,Keshavarzi:2019abf}.} Compared to~\cite{Passera:2008jk}, these have resulted in the following improvements to the data used in this analysis:
\begin{itemize}
\item In~\cite{Passera:2008jk}, below 500 MeV, a chiral perturbation theory (ChPT) estimation of the pion form factor $F_\pi(s)$ was used to estimate the hadronic cross section and overcome the lack of precise data in this region. In~\cite{Keshavarzi:2019abf}, measured data for all contributing modes in this region are included down to $\sim 300$ MeV, with ChPT used below this energy.
\item In the region $0.5 \leq \sqrt{s} \leq 1.4$ GeV where exclusive hadronic modes were used in~\cite{Passera:2008jk}, eight leading modes were summed to estimate the total cross section. The KNT19 data~\cite{Keshavarzi:2019abf} sums combined data for $> 33$ hadronic modes from $m_{\pi^0} \leq \sqrt{s} \lesssim 2$ GeV. It is estimated that $< 0.1\%$ of all contributions to $a_{\mu}^{\rm had,\,VP}$ remain missing from $\sigma_{\rm had} (s)$~\cite{Davier:2017zfy}.
\item Previously, from $1.4 \leq \sqrt{s} \leq 2$ GeV, inclusive measurements of the total hadronic cross section were used. Compared to the available exclusive measurements at these energies, these data are of poor quality with large errors. In~\cite{Keshavarzi:2019abf}, all available exclusive data are utilized to determine the total cross section.
\end{itemize}
\begin{table}[!t]
\vspace{-0.cm}
\centering
\scalebox{1.0}{
{\renewcommand{\arraystretch}{1.3}
\begin{tabular}{|c|c|c|}
\hline 
Observables & HMNT07~\cite{Hagiwara:2006jt} & KNT19 ~\cite{Keshavarzi:2019abf} \\
\hline 
$a_{\mu}^{\rm had,\,LO\,VP} \times 10^{10}$ & $ 698.4(4.8)$ & $ 692.8(2.4) $ \\ 
$\Delta\alpha_{\rm had}^{(5)}(M_{Z}^2) \times 10^{4}$ & $ 276.8 (2.2) $ & $ 276.1 (1.1) $ \\
\hline 
\end{tabular} 
}
}\caption{Comparison of the values of $a_{\mu}^{\rm had,\,LO\,VP}$ and $\Delta\alpha_{\rm had}^{(5)}(M_{Z}^2)$ used in~\cite{Passera:2008jk} and their corresponding values from the data used in this work.}\label{tab:KNTcompare}
\end{table}
These improvements can be summarized by comparing the values of $a_{\mu}^{\rm had,\,LO\,VP}$ and $\Delta\alpha_{\rm had}^{(5)}(M_{Z}^2)$ used in~\cite{Passera:2008jk} and their updated values employed in this work, which are given in Table~\ref{tab:KNTcompare}. In each case, the error has halved.

The increase in new hadronic cross section data has also resulted in more robust uncertainties. These, in particular, have been improved by the inclusion of the data covariances. In many cases, the experiments either provide information on the nature of the statistical and systematic correlations between data points or provide covariance matrices with the data. The KNT19 $\sigma_{\rm had} (s)$ data are complete with a full covariance matrix over the entire energy range. This has allowed for the error estimates in this work to be determined including all correlation information, which was not available in~\cite{Passera:2008jk}.

\subsection{Updates to EW precision measurements and their global fit}\label{sec:EWmeasurements}

The focus of the previous analysis~\cite{Passera:2008jk} was to assess the impact of the muon $g$-2 discrepancy on the prediction of the mass of the SM Higgs boson, $M_H$, which had not yet been discovered. Although the global fit to the EW data employs a large set of observables, the SM bounds on $M_H$ are strongly driven by the comparison of the theoretically accurate predictions for the $W$ boson mass $M_W$ and the effective EW mixing angle $\sin^2 \! \theta^{\rm lep}_{\rm eff}$ with their precisely measured values. For this reason, the analysis of~\cite{Passera:2008jk} employed simple analytic formulae providing the precise SM prediction of $M_W$ and $\sin^2 \! \theta^{\rm lep}_{\rm eff}$ in terms of $M_H$, $\Delta\alpha_{\rm had}^{(5)}(M_Z^2)$, the top quark mass $m_t$, and the strong coupling constant at the scale $M_Z$, $\alpha_s(M_Z^2)$~\cite{Degrassi:1997iy,Degrassi:1999jd,Ferroglia:2002rg,Awramik:2003rn,Awramik:2004ge,Awramik:2006uz,Degrassi:2014sxa}. 

A motivating factor for this work comes from the improved measurements of these EW observables. The values of all the parameters used in~\cite{Passera:2008jk} and their updated values used in this work are compared in Table~\ref{tab:EWcompare}. They are in good agreement with reduced uncertainties for all measurements.
These improvements provide additional constraints on the impact of the muon $g$-2 discrepancy on the EW sector of the SM. The most notable of these is the discovery, confirmation and subsequent detailed measurements of the Higgs boson~\cite{Aad:2012tfa,Chatrchyan:2012xdj,Aad:2015zhl,Aaboud:2018wps,CMS:2019drq}. The present PDG value $M_W=80.379(12)$~GeV, which is a combination of the ATLAS measurement with the LEP and Tevatron averages~\cite{PDG2018}, is lower than the value employed in~\cite{Passera:2008jk}. The value $\sin^2 \! \theta^{\rm lep}_{\rm eff}=0.23151(14)$, used in this work, is the combination of all $Z$-pole measurements performed at lepton (LEP and SLC) and hadron (Tevatron and LHC) colliders~\cite{Erler:2019hds}.
\begin{table}[!t]
\vspace{-0.cm}
\centering
\scalebox{1.0}{
{\renewcommand{\arraystretch}{1.3}
\begin{tabular}{|c|c|c|}
\hline 
EW parameter & Value used in~\cite{Passera:2008jk} (2008) & Value used in this work (2020) \\
\hline 
$M_W$ [GeV] & $ 80.398(25) $~\cite{Alcaraz:2006mx,Abazov:2003sv,Grunewald:2008zz} & $ 80.379(12) $~\cite{PDG2018} \\
$m_t$ [GeV] & $ 172.6 (1.4) $~\cite{Group:2008nq} & $ 172.9(4) $~\cite{PDG2018,TevatronElectroweakWorkingGroup:2016lid,Aaboud:2018zbu,Sirunyan:2018mlv} \\ 
$\alpha_s(M_Z^2)$ & $ 0.118(2) $~\cite{PDG2006} & $ 0.1179(10) $~\cite{PDG2018} \\
$\Delta\alpha_{\rm had}^{(5)}(M_Z^2)$ & $ 0.02768(22) $~\cite{Hagiwara:2006jt} & $ 0.02761(11) $~\cite{Keshavarzi:2019abf} \\
$M_H$ [GeV] & -- & $ 125.10(14) $~\cite{PDG2018,Aad:2015zhl,Aaboud:2018wps,CMS:2019drq} \\ 
$\sin^2 \! \theta^{\rm lep}_{\rm eff}$ & $ 0.23153(16) $~\cite{Alcaraz:2007ri} & $ 0.23151(14) $~\cite{Erler:2019hds} \\
\hline 
\end{tabular} 
}
}\caption{Comparison of the values of the EW precision observables used in~\cite{Passera:2008jk} and in this work.}
\label{tab:EWcompare}
\end{table}

The results derived from the simple analytic formulae for $M_W$ and $\sin^2 \! \theta^{\rm lep}_{\rm eff}$~\cite{Degrassi:1997iy,Degrassi:1999jd,Ferroglia:2002rg,Awramik:2003rn,Awramik:2004ge,Awramik:2006uz,Degrassi:2014sxa}
agree well with those of the global EW fit obtained using the publicly available package Gfitter~\cite{Flacher:2008zq,Baak:2011ze,Baak:2012kk,Baak:2014ora,Haller:2018nnx}. For example, the $M_H$ values predicted by the simple formulae of~\cite{Awramik:2003rn,Awramik:2004ge,Awramik:2006uz} and by Gfitter, using the latest inputs of Tables~\ref{tab:EWcompare} and~\ref{tab:GfitterResult}, are 
$97^{\,+20}_{\,-17}$~GeV
and
$94^{\,+20}_{\,-18}$~GeV,
respectively. In spite of this agreement and the simplicity of the analytic formulae, the main results presented in this work were obtained using the Gfitter package, which is more comprehensive with respect to the many EW observables that enter the global fit. Wherever possible, the Gfitter results presented here were checked against those obtained using the simple analytic formulae discussed above.

In predicting a specific parameter, the Gfitter package can either use the measured value of that parameter to constrain the fit result, or the experimental input can be omitted from the fit. In the latter case, the parameter is allowed to float freely, and the result is then determined from the fit of all other input parameters. Using the most recent values for the measurements of the EW parameters, the results obtained here using the Gfitter package are displayed in Table~\ref{tab:GfitterResult}. For the global `Fit result' given in the third column, all experimental inputs were used to constrain the fit. Good agreement is observed between these results and those given in the most recent Gfitter analysis update~\cite{Haller:2018nnx}. As the specific focus of this work is to predict the values of $M_W$, $\sin^2 \! \theta^{\rm lep}_{\rm eff}$, $M_H$ and $\Delta\alpha_{\rm had}^{(5)}(M_Z^2)$, fits excluding experimental inputs are only performed for those parameters. These results are displayed in the fourth column of Table~\ref{tab:GfitterResult} and agree well with the values $M_W = 80.354(7)$, $\sin^2 \! \theta^{\rm lep}_{\rm eff} = 0.23153(6)$, $M_H = 90^{\,+21}_{\,-18}$~GeV and $\Delta\alpha_{\rm had}^{(5)}(M_Z^2) = 271.6(3.9) \times 10^{-4}$ presented in~\cite{Haller:2018nnx}. The uncertainties of the fit results in the third column of Table~\ref{tab:GfitterResult} are parametric, i.e.\ due to the experimental errors of the inputs. As such, they do not include theoretical uncertainties, which are mainly due to unknown higher-order corrections and ambiguities in the definition of the top quark mass. The parametric uncertainties of the $M_W$, $\sin^2 \! \theta^{\rm lep}_{\rm eff}$, $M_H$ and $\Delta\alpha_{\rm had}^{(5)}(M_Z^2)$ predictions can be combined with their respective theoretical uncertainties, which have been estimated to be 
$\delta M_W = 5$~MeV, 
$\delta\sin^2 \! \theta^{\rm lep}_{\rm eff} = 4.3 \times 10^{-5}$, 
$\delta M_H = 6$~GeV, and 
$\delta \Delta\alpha_{\rm had}^{(5)}(M_Z^2) = 1.2 \times 10^{-4}$~\cite{Haller:2018nnx}. It should be noted that, even adding their theoretical uncertainties to the parametric ones, the fit predictions for $M_W$ and $\sin^2 \! \theta^{\rm lep}_{\rm eff}$ are more precise than their respective measured values. 

\begin{table}[!t]
\vspace{-0.cm}
\centering
\scalebox{1.0}{
{\renewcommand{\arraystretch}{1.3}
\hspace{-0.45cm}
\begin{tabular}{|c|c c|c|c|} 
\hline
Parameter & Input value & Reference & Fit result & Result w/o input value \\
\hline
$M_{W}$ [GeV] & $80.379(12)$ & \cite{PDG2018} & $80.359(3)$ & $80.357(4)(5)$ \\
$M_{H}$ [GeV] & $125.10(14)$ & \cite{PDG2018} & $125.10(14)$ & $94^{\,+20 ~ +6}_{\,-18 ~ -6}$ \\
$\Delta\alpha_{\rm had}^{(5)}(M_Z^2)\times 10^4$ & $276.1 (1.1)$ & \cite{Keshavarzi:2019abf} & $275.8 (1.1)$ & $272.2(3.9)(1.2)$ \\
$m_{t}$ [GeV] & $172.9(4)$ & \cite{PDG2018}& $173.0(4)$ & -- \\
$\alpha_{s}(M_{Z}^{2})$ & $0.1179(10)$ & \cite{PDG2018} & $0.1180(7)$ & -- \\
$M_{Z}$ [GeV] & $91.1876(21)$ & \cite{PDG2018} & $91.1883(20)$& -- \\
$\Gamma_{Z}$ [GeV] & $2.4952(23)$ & \cite{PDG2018} & 2.4940(4) & -- \\
$\Gamma_{W}$ [GeV] & $2.085(42)$ & \cite{PDG2018} & 2.0903(4) & -- \\
$\sigma_{\rm had}^{0}$ [nb] & $41.541(37)$ & \cite{ALEPH:2005ab} & 41.490(4) & -- \\
$R^{0}_{\l}$ & $20.767(25)$ & \cite{ALEPH:2005ab} & 20.732(4) & -- \\
$R^{0}_{c}$ & $0.1721(30)$ & \cite{ALEPH:2005ab} & 0.17222(8) & -- \\
$R^{0}_{b}$ & $0.21629(66)$ & \cite{ALEPH:2005ab} & $0.21581(8)$ & -- \\
$\bar{m_c}$ [GeV] & $1.27(2)$ & \cite{PDG2018} & $1.27(2)$ & -- \\
$\bar{m_b}$ [GeV] & $4.18^{\,+0.03}_{\,-0.02}$ & \cite{PDG2018} & $4.18^{\,+0.03}_{\,-0.02}$ & -- \\
\hdashline
$A_{\rm FB}^{0,\l}$ & $0.0171(10)$ & \cite{ALEPH:2005ab} & 0.01622(7) & -- \\
$A_{\rm FB}^{0,c}$ & $0.0707(35)$ & \cite{ALEPH:2005ab} & 0.0737(2) & -- \\
$A_{\rm FB}^{0,b}$ & $0.0992(16)$ & \cite{ALEPH:2005ab} & 0.1031(2) & -- \\$A_\ell$ & $0.1499(18)$ & \cite{Haller:2018nnx,ALEPH:2005ab} & 0.1471(3) & -- \\
$A_{c}$ & $0.670(27)$ & \cite{ALEPH:2005ab} & 0.6679(2) & -- \\
$A_{b}$ & $0.923(20)$ & \cite{ALEPH:2005ab} & 0.93462(7) & -- \\
$\sin^2 \! \theta^{\rm lep}_{\rm eff}(Q_{\rm FB})$ & $0.2324(12)$ & \cite{ALEPH:2005ab} & $0.23152(4)$ & $0.23152(4)(4)$ \\
$\sin^2 \! \theta^{\rm lep}_{\rm eff}(\rm Had. Coll.)$ & $0.23140(23)$ & \cite{Erler:2019hds} & $0.23152(4)$ & $0.23152(4)(4)$ \\
\hline
\end{tabular}
}
}\caption{Input values and fit results for the parameters entering the Gfitter global EW fit~\cite{Flacher:2008zq,Baak:2011ze,Baak:2012kk,Baak:2014ora,Haller:2018nnx}. For $M_W$, $\sin^2 \! \theta^{\rm lep}_{\rm eff}$, $M_H$ and $\Delta\alpha_{\rm had}^{(5)}(M_Z^2)$, which are the focus of this work, the last column gives the values for these observables predicted by the fit when their measured values are not included as input parameters. The errors quoted in the third column do not include theoretical uncertainties, while the errors in the fourth column are the experimental and theoretical uncertainties, respectively. The bottom eight rows, separated by a dashed line, indicate single measurements used by Gfitter which lead to the $\sin^2 \! \theta^{\rm lep}_{\rm eff}$ experimental average from $Z$-pole measurements.}
\label{tab:GfitterResult}
\end{table}

Gfitter uses separate inputs for single measurements which lead to the present $\sin^2 \! \theta^{\rm lep}_{\rm eff}$ experimental average from $Z$-pole measurements. They are listed in the eight bottom lines of Table~\ref{tab:GfitterResult}, separated by a horizontal dashed line. They include the forward-backward asymmetries and asymmetry parameters measured at LEP and SLC, the value of the effective EW mixing angle $\sin^2 \! \theta^{\rm lep}_{\rm eff}(Q_{\rm FB})$ derived from the forward-backward charge asymmetry measurement in inclusive hadronic events at LEP, and the combination $\sin^2 \! \theta^{\rm lep}_{\rm eff}(\rm Had. Coll.)$ of the LHC and Tevatron measurements. If all of these eight measurements (some of which are inconsistent with one another) are ignored, Gfitter's prediction for $M_H$ is 
$82^{\,+22}_{\,-19}$~GeV.
This value, driven by the $W$ boson mass, is lower than Gfitter's standard output 
$94^{\,+20}_{\,-18}$~GeV 
obtained employing all inputs (except $M_H$),
thereby increasing the tension with the $M_H$ measured value. All of these eight measurements are ignored when Gfitter is employed to predict the value of $\sin^2 \! \theta^{\rm lep}_{\rm eff}$ omitting its experimental value (last column of Table~\ref{tab:GfitterResult}).

The value $\Delta\alpha_{\rm had}^{(5)}(M_{Z}^2) = 0.02722(41)$ predicted from the EW fit, which yields the 95\%CL bounds $0.02642<\Delta\alpha_{\rm had}^{(5)}(M_{Z}^2)<0.02802$, agrees well with the dispersion relation result \allowbreak $\Delta\alpha_{\rm had}^{(5)}(M_{Z}^2)[{\rm KNT19}] = 0.02761(11)$. Although the uncertainty from the EW fit is too large to make significant statements, the smaller mean value is suggestive of a larger discrepancy than the current $\Delta a_{\mu}$. A less inclusive constraint on $\Delta\alpha_{\rm had}^{(5)}(M_Z^2)$ can be obtained using the natural relation that compares $\alpha$, the Fermi constant $G_F$, $M_W$ and $\sin^2 \! \theta^{\rm lep}_{\rm eff}$ at the level of their radiative corrections~\cite{Sirlin:1980nh,Marciano:1980pb,Marciano:1980be,Marciano:2004hb}, 
\beq \label{eq:natural}
\frac{\pi \alpha}{\sqrt{2} G_F M_W^2\sin^2 \! \theta^{\rm lep}_{\rm eff}} = 1-RC,
\eeq
where $1-RC = 0.9568(2) - \Delta \alpha_{\rm had}^{(5)}(M_Z^2)$. That combination of precision measurements is chosen because $RC$ is relatively insensitive to $M_H$ as well as various types of new physics effects. Using the experimental values given in Table~\ref{tab:EWcompare}, one finds $\Delta \alpha_{\rm had}^{(5)}(M_Z^2) = 0.02761(66)$. The agreement between this and the KNT19 dispersion relation result $\Delta \alpha_{\rm had}^{(5)}(M_Z^2)[{\rm KNT19}] = 0.02761(11)$ is excellent, but the error is less constraining than the EW fit bound. It does, however, provide further reassurance in the dispersive result. Constraints obtained using $M_W^2 (1- M_W^2 / M_Z^2)$ as input rather than the $M_W^2 \sin^2 \! \theta^{\rm lep}_{\rm eff}$ product, along with the appropriate change in $RC$, lead to smaller $\Delta \alpha_{\rm had}^{(5)}(M_Z^2)$ values (with errors similar to those obtained from equation~\eqref{eq:natural}), which is also suggestive of a larger discrepancy $\Delta a_{\mu}$.

\section{\boldmath Shifting $\sigma_{\rm had} (s)$ to cancel the muon $g$-2 discrepancy} \label{sec:shiftingSigma}

\begin{figure}[!t] 
\centering
\includegraphics[width=0.7\textwidth]{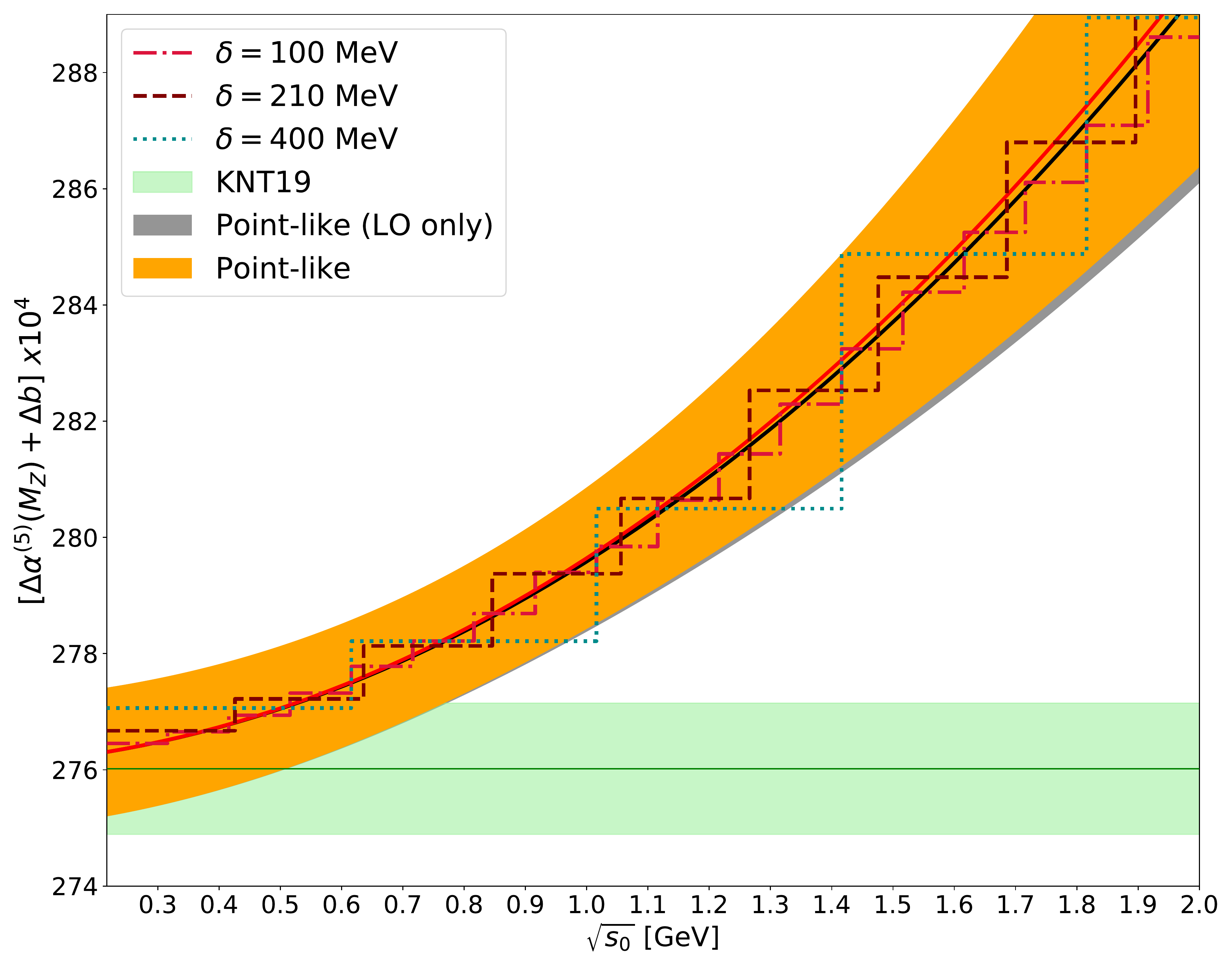}
\caption{\small Energy-dependent increases $\Delta\alpha_{\rm had}^{(5)}(M_{Z}^2) + \Delta b$ obtained when adjusting $\sigma_{\rm had}(s)$ to account for $\Delta a_{\mu}$. The dashed-crimson, dashed-maroon and dashed-cyan lines represent the binned shifts $\Delta b(\sqrt{s_0},\delta)$ for $\delta = 100,\,210,\,400$ MeV, respectively. The solid-red line displays the curve obtained for point-like increases $\Delta b(\sqrt{s_0})$, with the uncertainty given by the orange band. For a comparison of the result obtained in~\cite{Passera:2008jk}, the solid-black line and gray band show the result obtained when $f(s)$ accounts for only the LO contributions to $a_{\mu}^{\rm had,\,VP}$. The KNT19 result for $\Delta\alpha_{\rm had}^{(5)}(M_{Z}^2)$ is given by the light-green band~\cite{Keshavarzi:2019abf}.} \label{fig:dAlphaShift}
\end{figure}

The results for $\Delta\alpha_{\rm had}^{(5)}(M_{Z}^2) + \Delta b$ for both point-like and binned shifts of $\sigma_{\rm had} (s)$ (according to the procedure described in Section~\ref{sec:dAlphaShiftMethod}) are shown in Figure~\ref{fig:dAlphaShift}. The binned shifts have been computed for bin sizes of $\delta = 100, \, 210, \, 400$~MeV, all of which agree well with the point-like shifts defined by the solid red line and orange uncertainty band. The improvement from the inclusion of the dominant NLO contributions to $a_{\mu}^{\rm had,\,VP}$ is highlighted by the comparison of the solid red line with the solid black line, which shows the evaluation of the shifts $\Delta b$ with $f(s)$ only containing the LO contributions. As the NLO term provides a negative addition to $a_{\mu}^{\rm had,\,VP}$, the solid-red line is higher than the solid-black line. The change, however, is small, indicating that the results presented in~\cite{Passera:2008jk}, which did not include the NLO contributions, are not adversely affected by their omission.

\subsection{Impact on $M_W$, $\sin^2 \! \theta^{\rm lep}_{\rm eff}$ and $M_H$}\label{sec:Results}

\begin{figure}[!t] 
\centering
\subfloat[The $W$ boson mass, $M_W$.]{%
\includegraphics[width= 0.49\textwidth]{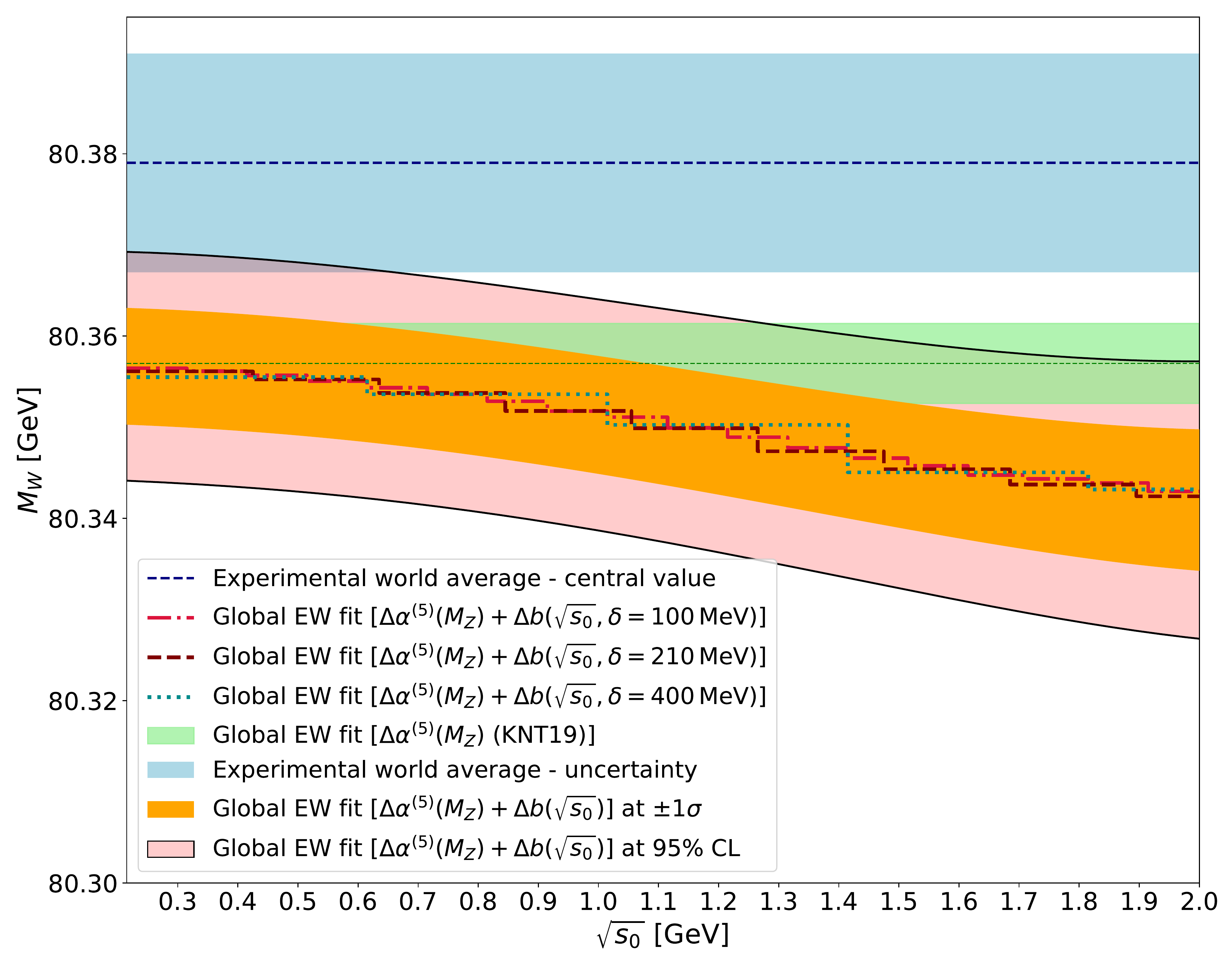}\label{fig:EWvarMW}}\hfill
\subfloat[The effective weak mixing angle, $\sin^2 \! \theta^{\rm lep}_{\rm eff}$.]{%
\includegraphics[width= 0.49\textwidth]{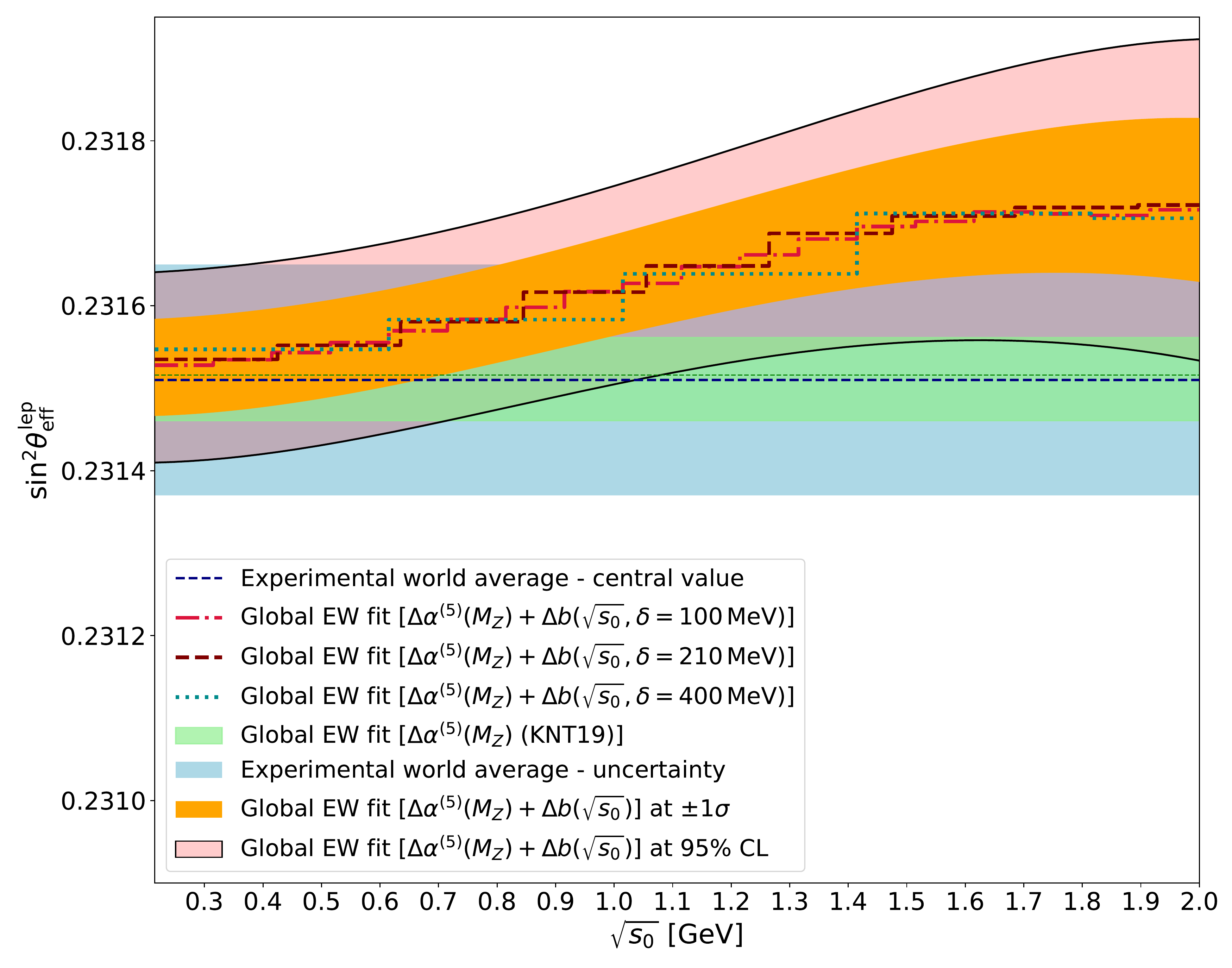}\label{fig:EWvarST}}\hfill
\subfloat[The Higgs boson mass, $M_H$.]{%
\includegraphics[width= 0.65\textwidth]{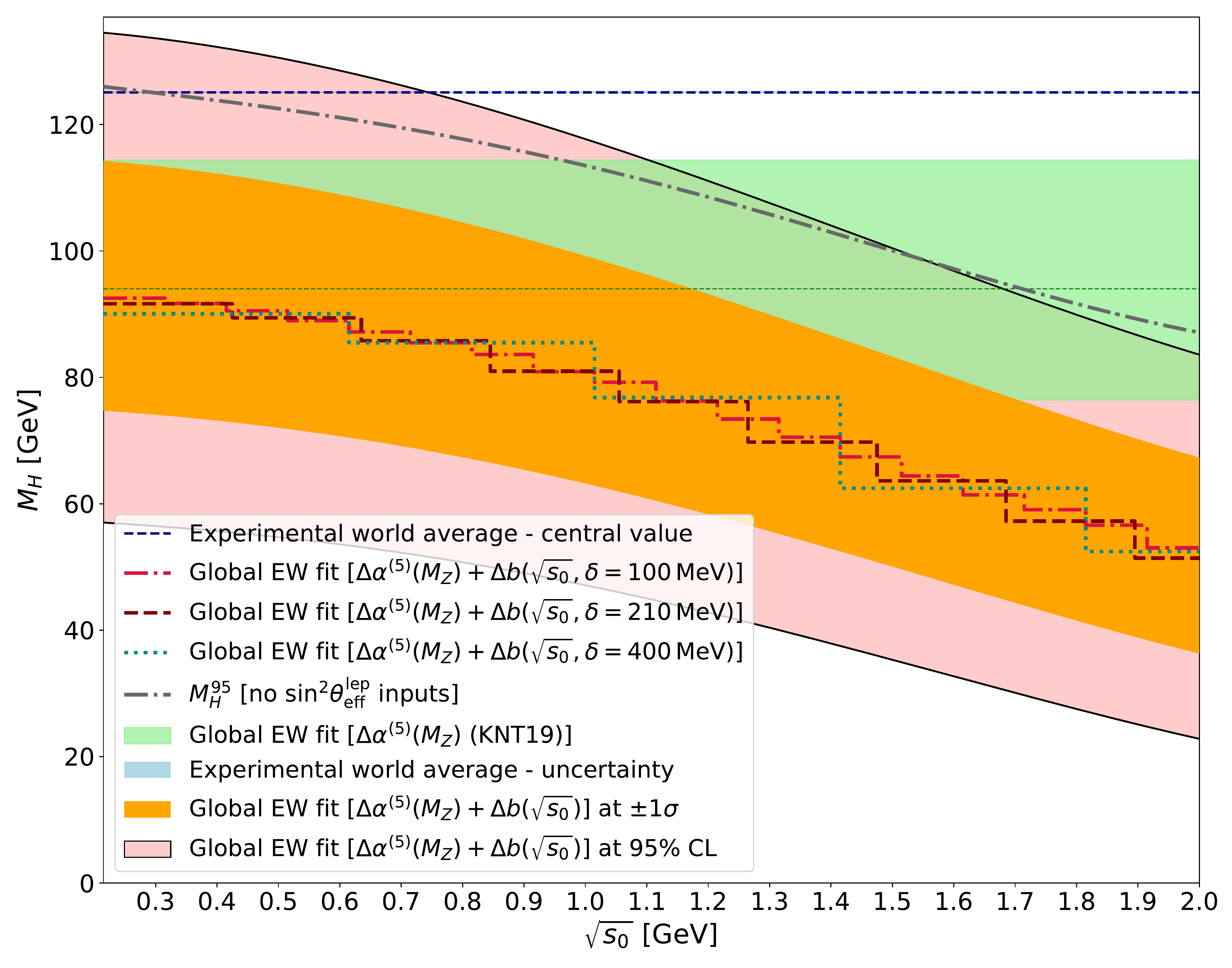}\label{fig:EWvarMH}}\hfill
\caption{\small The values of $M_W$, $\sin^2 \! \theta^{\rm lep}_{\rm eff}$ and $M_H$ obtained from the Gfitter global EW fit using as input the increased values of $\Delta\alpha_{\rm had}^{(5)}(M_{Z}^2) + \Delta b$ shown in Figure~\ref{fig:dAlphaShift}. The orange band displays the curve obtained for point-like increases $\Delta b(\sqrt{s_0})$, where the width of the band is the $\pm1\sigma$ uncertainty obtained from the fit (including both parametric and theoretical errors). The solid-black lines outlining the pink band indicate the upper/lower bounds of the EW observable determined from the fit at the 95\% CL. The dashed-crimson, dashed-maroon and dashed-cyan lines represent the value obtained from binned shifts $\Delta b(\sqrt{s_0},\delta)$ for $\delta = 100,\,210,\,400$ MeV, respectively. The result for each observable determined from the EW fit using the KNT19 result for $\Delta\alpha_{\rm had}^{(5)}(M_{Z}^2)$ is given by the light-green band. The dashed-navy line indicates the central value of the measured value of each EW observable, the uncertainty of which is displayed by the light-blue band. The gray dash-dotted line in Figure~\ref{fig:EWvarMH} indicates Gfitter's prediction for the 95\%CL upper bound on $M_H$ obtained omitting all the measurements which lead to the present $\sin^2 \! \theta^{\rm lep}_{\rm eff}$ experimental average from $Z$-pole measurements (see Section~\ref{sec:EWmeasurements}).} \label{fig:EWvar}
\end{figure}

For each point-like and binned shift of $\sigma_{\rm had} (s)$ shown in Figure~\ref{fig:dAlphaShift}, the corresponding value of $\Delta\alpha_{\rm had}^{(5)}(M_{Z}^2) + \Delta b$ was used as input to determine $M_W$, $\sin^2 \! \theta^{\rm lep}_{\rm eff}$ and $M_H$ from the Gfitter package.\footnote{All results have been additionally checked against the simple analytic formulae discussed in Section~\ref{sec:EWmeasurements}.} For each evaluation, only the observable being determined was allowed to float freely in the fit, with all other parameters entering the fit with their experimentally measured values and uncertainties (except for $\Delta\alpha_{\rm had}^{(5)}(M_{Z}^2)$, which was shifted by $\Delta b$). The results of these variations for $M_W$, $\sin^2 \! \theta^{\rm lep}_{\rm eff}$ and $M_H$ are shown in Figure~\ref{fig:EWvar}.

The results for $M_W$, shown in Figure~\ref{fig:EWvarMW}, are driven by the standard prediction from the global EW fit, $M_W = 80.357(6)$ GeV (indicated by the light-green band), which is $1.6\sigma$ below the measured value of $M_W^{\rm exp} = 80.379(12)$ GeV~\cite{PDG2018}. This difference is further emphasized by the precision of $M_W$ which, as previously stated, is higher than that of $M_W^{\rm exp}$. An increase of the hadronic cross section to account for $\Delta a_\mu$ increases $\Delta\alpha_{\rm had}^{(5)}(M_{Z}^2)$. Figure~\ref{fig:EWvarMW} shows that this consequently reduces the predicted value of $M_W$, therefore increasing the difference with the measured value $M_W^{\rm exp}$. There is overlap between $M_W^{\rm exp}$ and the upper bound of the $M_W$ prediction at the 95\%CL for increases of the cross section at low energies, but accounting for $\Delta a_\mu$ in the hadronic vacuum polarization contributions is found here to be excluded for $\sqrt{s_0} \gtrsim 0.9$~GeV at the 95\%CL. For $\sqrt{s_0} = 2$ GeV, $M_W$ is $2.6\sigma$ below $M_W^{\rm exp}$. 

As with $M_W$, the predictions for the effective weak mixing angle from the EW fit are more precise than the experimental average of $\sin^2 \! \theta^{\rm lep, \, exp}_{\rm eff} = 0.23151(14)$~\cite{Erler:2019hds} shown in Figure~\ref{fig:EWvarST}. However, contrary to $M_W$, $\sin^2 \! \theta^{\rm lep}_{\rm eff}$ increases for larger values of $\Delta\alpha_{\rm had}^{(5)}(M_{Z}^2)$. The standard EW fit prediction of $\sin^2 \! \theta^{\rm lep}_{\rm eff} = 0.23152(6)$ GeV shown by the green band is in very good agreement with the measured value, and the increases due to $\Delta b$ are still consistent with the measured $\sin^2 \! \theta^{\rm lep, \, exp}_{\rm eff}$ for all $\sqrt{s_0}$ in the considered range. 

As found in~\cite{Passera:2008jk}, Figure~\ref{fig:EWvarMH} shows that increases in $\sigma_{\rm had} (s)$ and $\Delta\alpha_{\rm had}^{(5)}(M_{Z}^2)$, decrease the predicted value for $M_H$. The precise value of the measured Higgs mass is shown at $M_H^{\rm exp} = 125.10(14)$~GeV. In~\cite{Passera:2008jk}, without this measurement, the allowed parameter space was restricted by the LEP lower bound of $M_H^{\rm LB} \simeq 114.4$ GeV~\cite{Barate:2003sz} and hypothetical shifts in $\sigma_{\rm had} (s)$ to account for $\Delta a_\mu$ were excluded above $\sqrt{s_0} \gtrsim 1.1$ GeV for point-like shifts at the 95\%CL. From Figure~\ref{fig:EWvarMH}, shifts in the hadronic cross section needed to bridge the muon $g$-2 discrepancy via $a_{\mu}^{\rm had,\,VP}$ are found to be excluded for $\sqrt{s_0} \gtrsim 0.7$~GeV at the 95\%CL. Although this reduction in the allowed parameter space is significant in itself, importantly it now excludes the possibility of $\Delta a_\mu$ originating from a large portion of the $\rho$ (which is the single dominant resonant contribution to $a_{\mu}^{\rm had,\,VP}$) and $\omega$ resonances, plus the entire $\phi$ resonance above the strange quark threshold.

Figure~\ref{fig:EWvarMH} also shows Gfitter's prediction for the 95\%CL upper bound on $M_H$ obtained omitting all the measurements which lead to the present $\sin^2 \! \theta^{\rm lep}_{\rm eff}$ experimental average from $Z$-pole measurements, some of which are inconsistent with one another (see Section~\ref{sec:EWmeasurements}). For $\sqrt{s_0} \lesssim 1.6$~GeV, this upper bound, driven by the $W$ boson mass, is more stringent than the one obtained employing all inputs (except $M_H$).

\begin{figure}[!t] 
\centering
\subfloat[The $W$ boson mass, $M_W$.]{%
\includegraphics[width= 0.5\textwidth]{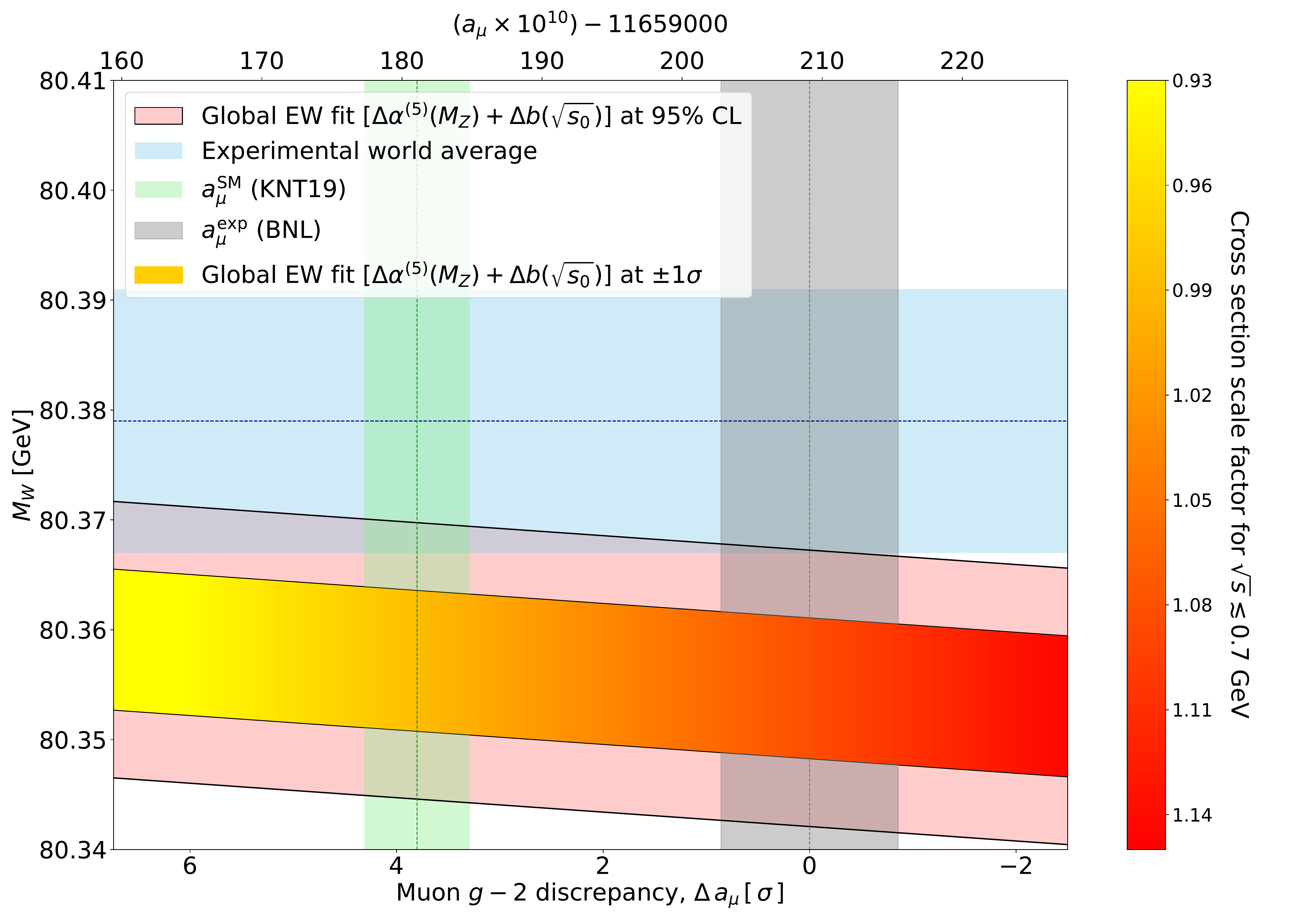}}\hfill
\subfloat[The effective weak mixing angle, $\sin^2 \! \theta^{\rm lep}_{\rm eff}$.]{%
\includegraphics[width= 0.5\textwidth]{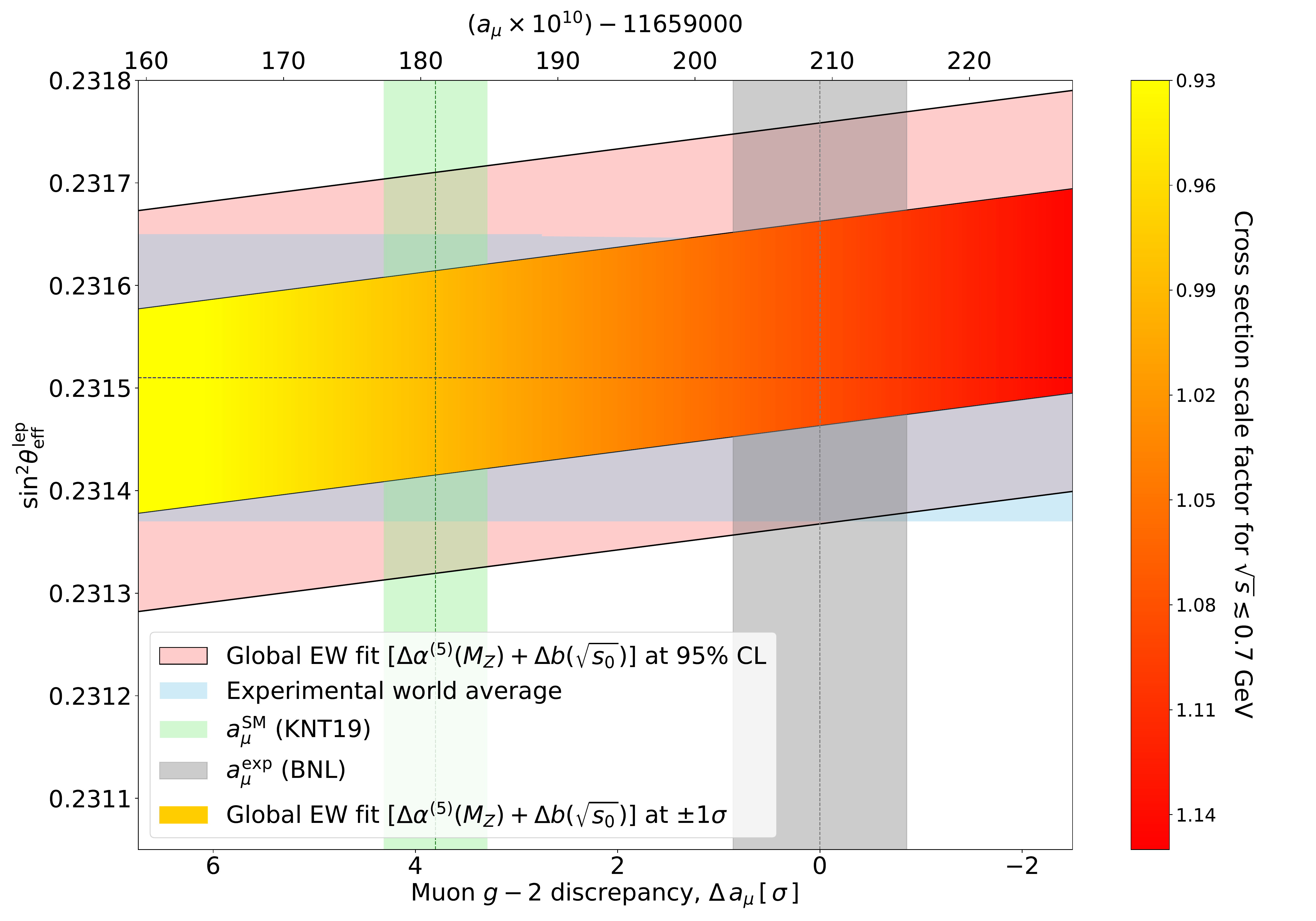}}\hfill
\subfloat[The Higgs boson mass, $M_H$.]{%
\includegraphics[width= 0.8\textwidth]{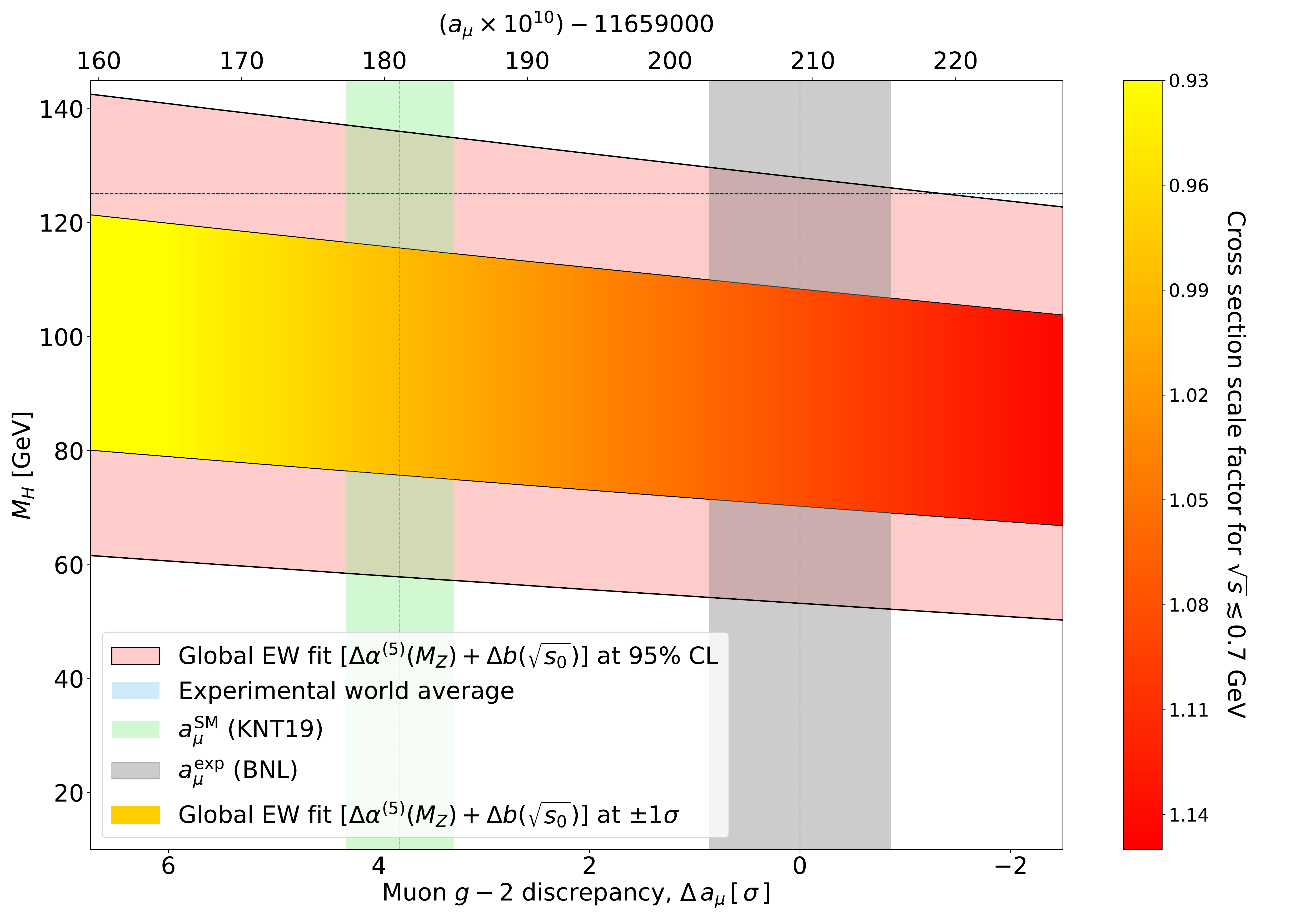}}\hfill
\caption{\small Plots of $M_W$, $\sin^2 \! \theta^{\rm lep}_{\rm eff}$ and $M_H$ vs. $a_\mu$ and the resulting muon $g$-2 discrepancy $\Delta a_{\mu}[\sigma]$ obtained when scaling $\sigma_{\rm had}(s)$ in a single energy-independent region $\sqrt{s} \lesssim 0.7$~GeV. In each case, the graded yellow-red band displays the value of each observable obtained from the Gfitter global EW fit, using as input the shifted values of $\Delta\alpha_{\rm had}^{(5)}(M_{Z}^2)$ determined from the scaled $\sigma_{\rm had}(s)$. The width of this band is the $\pm1\sigma$ uncertainty obtained from the fit (including both parametric and theoretical errors). The yellow-red color-gradient of this band and the z-axis indicates the magnitude of the scale factor $\epsilon$ that is applied to account for $\Delta a_\mu$. The solid-black lines outlining the pink band indicate the upper/lower bounds of the EW observable determined from the fit at the 95\% CL. The light-blue band indicates the measured value of each EW observable. The KNT19 result for $a_\mu^{\rm SM}$ is given by the light-green band, corresponding to $\Delta a_{\mu}=3.8\sigma$~\cite{Keshavarzi:2019abf}. The experimental measurement of $a_\mu$~\cite{Bennett:2002jb,Bennett:2004pv,Bennett:2006fi,PDG2018}, indicating a no-new-physics scenario and corresponding to $\Delta a_{\mu}=0$, is given by the gray band.} \label{fig:EWvsGm2}
\end{figure}

In an attempt to further scrutinize these findings, the predictions for $M_W$, $\sin^2 \! \theta^{\rm lep}_{\rm eff}$ and $M_H$ were obtained varying $\sigma_{\rm had}(s)$ by a multiplicative, percentage value $\epsilon$ in a single, energy-independent bin ranging from $m_{\pi^0}$ to $\sim0.7$~GeV. These results, shown in Figure~\ref{fig:EWvsGm2}, are as expected given the conclusions from Figure~\ref{fig:EWvar}. The difference between $M_W^{\rm exp}$ and the EW fit prediction increases when adjusting $\Delta\alpha_{\rm had}^{(5)}(M_{Z}^2)$. The point where the $a_\mu$ prediction coincides with $a_\mu^{\rm exp}$, which yields $\Delta\alpha_{\rm had}^{(5)}(M_{Z}^2) = 277.6(1.1) \times 10^{-4}$, corresponds to a difference of $1.8\sigma$ for $M_W$. However, there is overlap with the experimental uncertainty for the $M_W^{95}$ upper bound for both $a_\mu^{\rm SM}$ and $a_\mu^{\rm exp}$. In all cases considered, the predicted values of $\sin^2 \! \theta^{\rm lep}_{\rm eff}$ agree with the measured one. For the Higgs mass, there is also agreement between the upper bound $M_H^{95}$ from the EW fit and $M_H^{\rm exp}$ for $\Delta\alpha_{\rm had}^{(5)}(M_{Z}^2)$ corresponding to both $a_\mu^{\rm SM}$ and $a_\mu^{\rm exp}$. These results indicate that, at the 95\%CL, there is room in the EW fit to allow for such an increase to $\sigma_{\rm had}(s)$ below $\sim0.7$~GeV which would account for $\Delta a_\mu$. 

\begin{figure}[!t] 
\centering
\subfloat[Linear scale.]{%
\includegraphics[width= 0.8\textwidth]{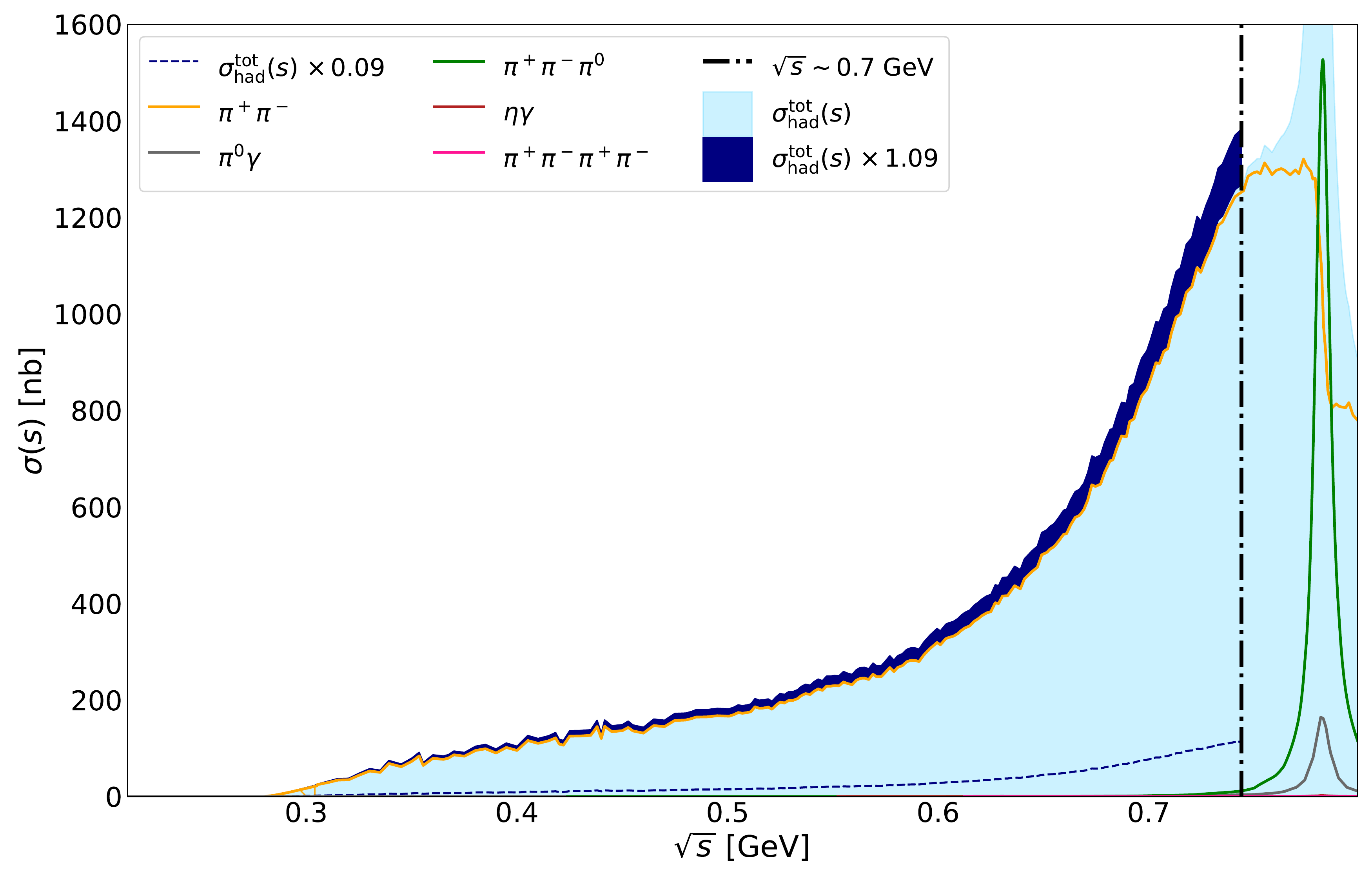}\label{fig:epsilon_norm}}\hfill
\subfloat[Log scale.]{%
\includegraphics[width= 0.8\textwidth]{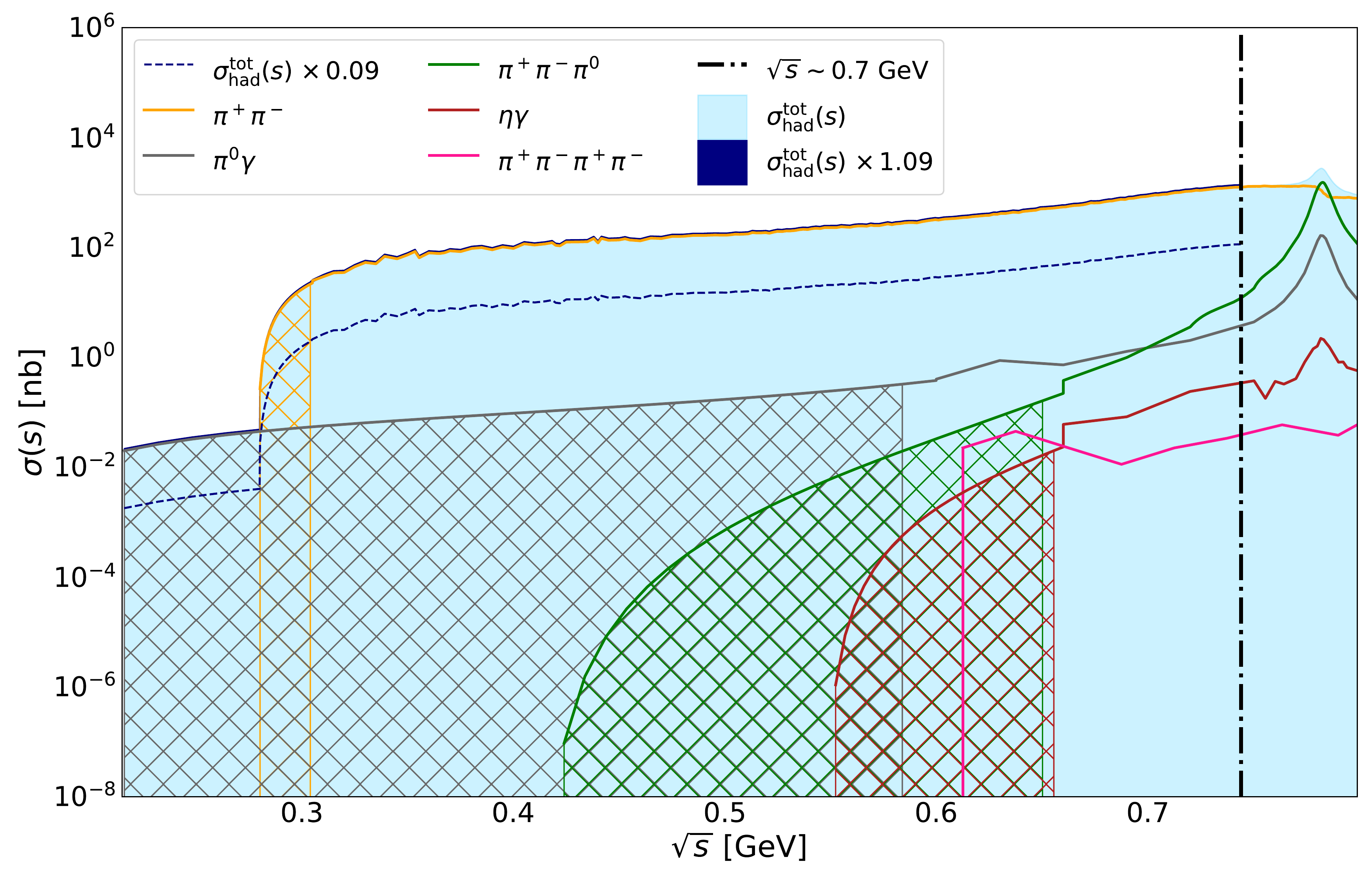}\label{fig:epsilon_log}}\hfill
\caption{\small Contributions to the total hadronic cross section from the hadronic modes contributing below $\sim0.7$~GeV, plotted for a linear y-axis (a) and for the y-axis represented on a log scale (b). The total cross section is shown by the filled light-blue region in each plot. The vertical dashed-black line indicates the exact point at $\sqrt{s}\sim0.7$~GeV below which the navy-blue dashed line displays the $\epsilon = +9\%$ shift of the cross section that is applied to account for $\Delta a_\mu$. The applied increase of $\epsilon = +9\%$ to the total cross section is given by the filled navy area. For each of the hadronic modes, the cross-hatch pattern indicates the portion of that hadronic mode that is estimated from ChPT (plus $\omega$-dominance for the $\pi^+\pi^-\pi^0$ and $\pi^0\gamma$ channels)~\cite{Hagiwara:2003da} due to a lack of experimental data at threshold.} \label{fig:epsilonChPT}
\end{figure}

The question that then arises is: should the discrepancy $\Delta a_\mu$ be bridged by an increase in the hadronic cross section from $m_{\pi^0}$ to $\sim 0.7$~GeV, is the required change $\epsilon$ realistic? The magnitude of $\epsilon$ required is depicted in Figure~\ref{fig:EWvsGm2} by the yellow-red color-gradient of the band showing the $\pm1\sigma$ uncertainty obtained from the EW fit. For the results corresponding to $a_\mu^{\rm exp}$, the factor required is $\epsilon \approx +9\%$. This adjustment is shown in Figure~\ref{fig:epsilonChPT}, which displays the hadronic final states determined in~\cite{Keshavarzi:2019abf} that contribute to the total hadronic cross section below $\sim0.7$~GeV. It is clear that the total cross section below this energy is entirely dominated by the $\pi^+\pi^-$ channel, where Figure~\ref{fig:epsilon_norm} highlights how the choice to make an energy-independent increase, even over a small energy region such as this, is not locally representative of the contributing final states. In this case, an $\epsilon = +9\%$ increase to the integrated cross section captures most of the leading tail of the $\rho$-resonance in the $\pi^+\pi^-$ channel. The result is that the increase to account for $\Delta a_\mu$ has come overwhelmingly from an increase to the $\rho$ as opposed to the other hadronic modes also present in this region. Given the high quantity of precise measurements of the $\pi^+\pi^-$ channel, especially on the $\rho$-resonance, it is highly unlikely that an integrated contribution of $9\%$ would be absent from the measured data.

Of interest, however, are the threshold contributions of $a_\mu^{\pi^+\pi^-}(\sqrt{s} \lesssim 0.3 \, {\rm GeV})$ , $a_\mu^{\pi^0\gamma}(\sqrt{s} \lesssim 0.6 \, {\rm GeV})$, $a_\mu^{\pi^+\pi^-\pi^0}(\sqrt{s} \lesssim 0.7 \, {\rm GeV})$ and $a_\mu^{\eta\gamma}(\sqrt{s} \lesssim 0.7 \, {\rm GeV})$ that are estimated from ChPT (plus $\omega$-dominance contributions for the $\pi^+\pi^-\pi^0$ and $\pi^0\gamma$ channels). The experimental data above these given energies covers the majority of each cross section below $\sim2$ GeV, with the largest respective ChPT estimate being in the $\pi^0\gamma$ channel, which accounts for $< 3\%$ of the total $a_\mu^{\pi^0\gamma}$. Nevertheless, the required level of precision of $a_{\mu}^{\rm had,\,VP}$ necessitates that theoretical estimates of the cross section, in this case from ChPT, are used to account for the lack of experimental data at the production thresholds of each of these modes. These are shown in Figure~\ref{fig:epsilon_log} by the cross-hatch pattern for each final state and cover a large portion of the region below $\sim0.7$~GeV.\footnote{For completeness, the measured $\pi^+\pi^-\pi^+\pi^-$ contribution below $\sim0.7$~GeV is also shown in Figure~\ref{fig:epsilonChPT}. Although these data begin at $\sim 610$ MeV with a non-zero value of $\sigma_{2\pi^+2\pi^-}\sim0.2$ nb~\cite{Lees:2012cr}, thereby causing the unfortunate step-function in the start of this measurement shown in Figure~\ref{fig:epsilon_log}, the production threshold of $m_{2\pi^+2\pi^-} \approx 558$~MeV leaves an unmeasured window of roughly 50~MeV, which is not a cause for concern.} 

The $\pi^0\gamma$ contribution is estimated over the widest energy interval. For such a wide region being estimated by theoretical predictions of the cross section, it is difficult to make firm statements about the possibility of missed contributions without corroborative experimental measurements. This is especially relevant when considering the comparison of the results from dispersive approaches with the recent LQCD result for $a_{\mu}^{\rm had,\,LO\,VP}$ from the BMW collaboration~\cite{Borsanyi:2020mff}. This result, as previously mentioned, results in a value of $a_{\mu}^{\rm SM}$ that agrees with $a_{\mu}^{\rm exp}$ and is therefore in tension with the dispersive evaluations. Unlike needing to individually account for all exclusive modes at low energies as in the dispersive approach, the results from the lattice evaluations (which calculate $a_{\mu}^{\rm had,\,VP}$ based on quark flavor) inclusively capture all contributions, including those that may potentially be missed by the measurements of the hadronic cross section. This, plus the knowledge that portions of the cross section below $\sim0.7$~GeV are estimated from theory models, could suggest that the result from~\cite{Borsanyi:2020mff} may have included contributions from this region that are potentially absent from the hadronic data.

\begin{figure}[!t] 
\centering
\includegraphics[width=0.8\textwidth]{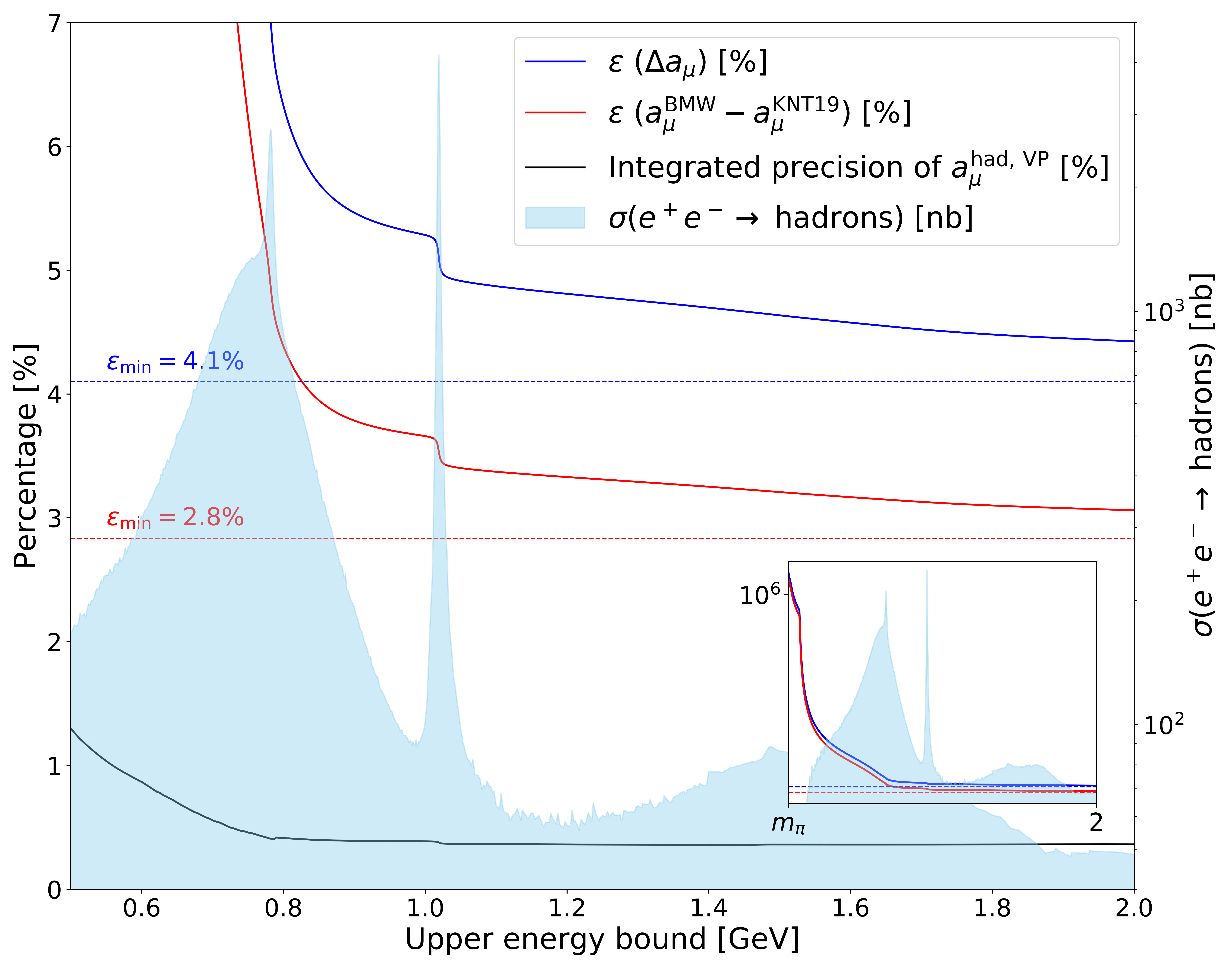}
\caption{\small The increase of $\epsilon$ (\%) required to account for the muon $g$-2 discrepancy (solid-blue line), as well as the discrepancy between the lattice BMW~\cite{Borsanyi:2020mff} and dispersive KNT19 results~\cite{Keshavarzi:2019abf} (solid-red line), when adjusting $\sigma_{\rm had} (s)$ from threshold up to a given upper energy bound. The dashed-blue and dashed-red lines show the minimum increase required (corresponding to an upper energy bound of $\infty$) for $\Delta a_\mu$ and the lattice BMW vs.\ dispersive KNT19 evaluations, respectively. The solid-black line depicts the precision in \% of $a_{\mu}^{\rm had,\,VP}$ when integrating from threshold to the upper energy bound. The total $e^+e^- \rightarrow \text{hadrons}$ cross section is given by the light-blue shaded region for reference.} \label{fig:epsilon}
\end{figure}
However, with the $\pi^0\gamma$ cross section estimated from ChPT resulting in a contribution of $a_{\mu}^{\pi^0\gamma}({\rm ChPT})= (0.12 \pm 0.01) \times 10^{-10}$ and, indeed, the contribution from all four ChPT estimated modes totaling $a_{\mu}^{\rm had, \, VP}({\rm ChPT})= (1.00\pm 0.02) \times 10^{-10}$, accounting for $\Delta a_{\mu} = (28.02 \pm 7.37)\times 10^{-10}$ would require an exceptionally large increase of the cross section relative to the already included theoretical prediction. Moreover, for each of the estimated modes, the transition points between the estimated cross section and the measured data are currently in good agreement for all four channels. Therefore, any additional hypothetical structure in the region from the respective production thresholds up to $\sim0.7$~GeV would have to be narrow enough to also agree with the measured data at these transition points. To put this into context, for the $\pi^0\gamma$ channel, the cross section in the ChPT-estimated region would have to exhibit an integrated increase of $\mathcal{O}(10^4)$ and decrease sharply to agree with $\sigma_{\pi^0\gamma}\sim 0.4$~nb where the data begins at 600 MeV. For the estimated $\pi^+\pi^-$ threshold contribution, the cross section would have to increase by $\mathcal{O}(10^3)$ over an extremely narrow width of roughly 25 MeV.

These considerations are summarized in Figure~\ref{fig:epsilon}. The solid-blue line shows the percentage increase $\epsilon$ required to account for the muon $g$-2 discrepancy when shifting the cross section in a bin spanning from the hadronic threshold to a given upper energy bound. Similarly, the discrepancy between the recent lattice BMW~\cite{Borsanyi:2020mff} and the dispersive KNT19~\cite{Keshavarzi:2019abf} results is shown by the solid-red line. As indicated by this plot, increasing the cross section (which is shown in Figure~\ref{fig:epsilon} in light-blue) in a purely energy-independent manner from $m_{\pi^0} \leq \sqrt{s} < \infty$ bounds the minimum increase required at $+4.1\%$ for $\Delta a_\mu$ and $+2.8\%$ for the lattice BMW vs.\ the dispersive KNT19 evaluations. Obviously, all values for $\epsilon$ increase when reducing this upper energy bound, culminating (as indicated in the full-range inset plot in Figure~\ref{fig:epsilon}) in the previously mentioned, orders of magnitude adjustment needed to account for $\Delta a_\mu$ in the ChPT-estimated region close to threshold. To argue the improbability that $\geq4.1\%$ (or even $\geq2.8\%$) contributions could have been missed in the many precise measurements that contribute to the total hadronic cross section and, indeed, to further highlight the relative magnitude of such a hypothetical increase, the integrated precision of $a_{\mu}^{\rm had,\,VP}$ is shown by the solid-black line. For the entire energy range, its precision is much smaller than all values of $\epsilon$, at roughly $\sim 0.5\%$ overall and increasing above 1\% only for the lowest energies where $\epsilon$ is irreconcilably large. Importantly, these values of $\epsilon$ are much larger than any current disagreements observed between hadronic cross section measurements from different experiments, the tensions of which are additionally accounted for in a $\chi^2$ error inflation procedure that contributes to the displayed precision of $a_{\mu}^{\rm had,\,VP}$~\cite{Hagiwara:2003da,Hagiwara:2006jt,Hagiwara:2011af,Keshavarzi:2018mgv,Keshavarzi:2019abf}.

\subsection{Prediction of $\Delta\alpha_{\rm had}^{(5)}(M_{Z}^2)$ from the EW fit and consequences for $a_\mu$} \label{sec:reverseArgument}

The prediction of $\Delta\alpha_{\rm had}^{(5)}(M_{Z}^2)$ from the EW fit is
\beq \label{delalphahad_EWfit}
\Delta\alpha_{\rm had}^{(5)}(M_{Z}^2) = 272.2(4.1) \times 10^{-4}\, , 
\eeq
where the error is the sum in quadrature of the experimental and theoretical uncertainties given in Table~\ref{tab:GfitterResult}. The lower mean value of equation~\eqref{delalphahad_EWfit} with respect to equation~\eqref{delalphahad_KNT19} is expected considering the EW fit predicts a lower value for $M_H$. This result is in good agreement with equation~\eqref{delalphahad_KNT19} (although far less precise) and the difference between them is 
\begin{align}
\Delta[\Delta\alpha_{\rm had}^{(5)}(M_{Z}^2) ] = -3.9(4.3) \times 10^{-4}. 
\end{align}
This presents an interesting opportunity to reverse the methodology described in Section~\ref{sec:dAlphaShiftMethod}. There, the case was made to adjust the hadronic cross section to account for $\Delta a_\mu$ and investigate what effect this had on the global EW fit. Here, the argument can therefore be made to adjust $\sigma_{\rm had}(s)$ to account for $\Delta[\Delta\alpha_{\rm had}^{(5)}(M_{Z}^2) ]$ predicted from the global EW fit and investigate what effect this has on $\Delta a_\mu$. As $\Delta[\Delta\alpha_{\rm had}^{(5)}(M_{Z}^2) ]$ is negative, the data for $\sigma_{\rm had}(s)$ should be decreased in a similar energy-dependent procedure as in Section~\ref{sec:dAlphaShiftMethod} based on binned and point-like shifts. $\Delta a_\mu$ is consequently modified by
\beq 
\Delta a_\mu' = \Delta a_\mu - \Delta a\, ,
\eeq
where $\Delta a$ is negative. 

\begin{figure}[!t] 
\centering
\includegraphics[width=0.8\textwidth]{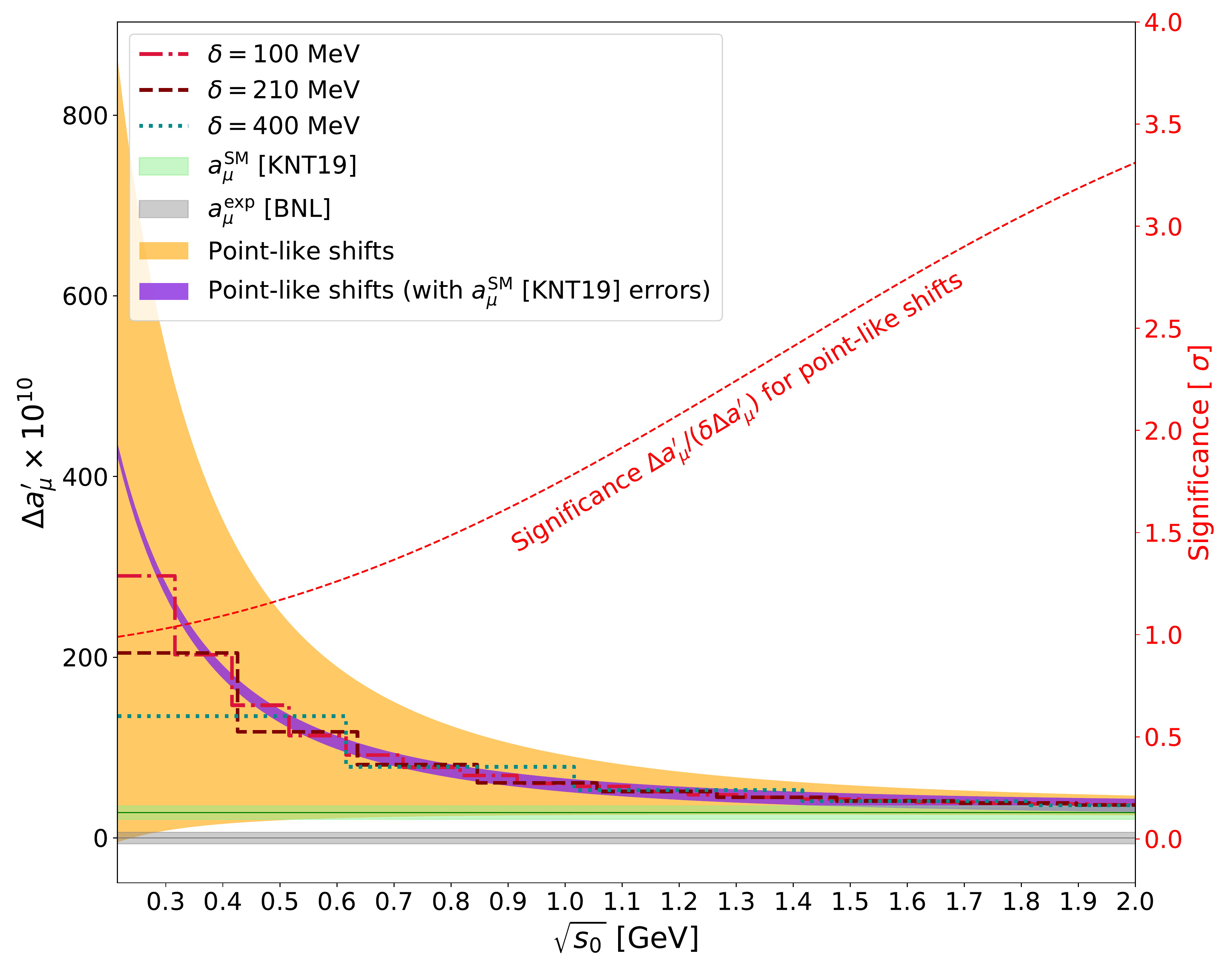}
\caption{\small Energy-dependent decreases $\Delta a_\mu'$ from adjustments to $\sigma_{\rm had}(s)$ to account for $\Delta[\Delta\alpha_{\rm had}^{(5)}(M_{Z}^2)] = -3.9(4.3) \times 10^{-4}$. The dashed-crimson, dashed-maroon and dashed-cyan lines represent the binned shifts for $\delta = 100,\,210,\,400$ MeV, respectively. The orange band defines the uncertainty for the curve obtained for point-like increases $\Delta b(\sqrt{s_0})$. The purple band shows the central values of $\Delta a_\mu'$, but with $\Delta a$ in each case forced to have the same precision as $a_\mu^{\rm SM}[{\rm KNT19}]$. The KNT19 result $\Delta a_\mu = (28.0 \pm 7.4)\times 10^{-10}$ is given by the light-green band~\cite{Keshavarzi:2019abf}. The experimental measurement of $a_\mu$~\cite{Bennett:2002jb,Bennett:2004pv,Bennett:2006fi,PDG2018}, corresponding to $\Delta a_{\mu}=0$, is given by the gray band. The dashed-red line, corresponding to the right-hand y-axis only, shows the significance in standard deviations of the difference between $a_\mu^{\rm exp}$ and the orange band.} \label{fig:reverseArgument}
\end{figure}

The corresponding results for $\Delta a_\mu'$ for both point-like and binned shifts of $\sigma_{\rm had} (s)$ are shown in Figure~\ref{fig:reverseArgument}. Again, the binned shifts for $\delta = 100, \, 210, \, 400$ MeV all agree well with the curve defining the point-like shifts, the uncertainty of which is given by the orange band. In all cases, adjusting $\sigma_{\rm had} (s)$ to account for $\Delta[\Delta\alpha_{\rm had}^{(5)}(M_{Z}^2)]$ results in an increase in the central value of $\Delta a_\mu$. In other words, the values for $\Delta a_\mu'$ obtained from the EW fit prediction of $\Delta\alpha_{\rm had}^{(5)}(M_{Z}^2)$ not only validate a positive difference between $a_\mu^{\rm exp}$ and $a_\mu^{\rm SM}$, but prefer a larger absolute difference than the current central value of $\Delta a_\mu$. However, these deviations are not as significant as $\Delta a_{\mu} = (28.0 \pm 7.4)\times 10^{-10}$ ($3.8\sigma$)~\cite{Keshavarzi:2019abf} due the large errors on $\Delta\alpha_{\rm had}^{(5)}(M_{Z}^2)$ arising from the EW fit. For point-like shifts in $\sigma_{\rm had} (s)$, the values for the resulting muon $g$-2 discrepancy range from $\Delta a_\mu' = (427.6 \pm 432.2) \times10^{-10}$, i.e.\ $1.0\sigma$, at the hadronic production threshold to $\Delta a_\mu' = (36.0 \pm 10.9) \times10^{-10}$, i.e.\ $3.3\sigma$, at 2 GeV. This is indicated by the dashed-red line in Figure~\ref{fig:reverseArgument}. Although this makes it difficult to draw conclusions based on statistical significance, it is possible to consider the hypothetical situation of the results for $\Delta a_\mu'$ corresponding to the central values of the orange band, but with uncertainties in $\Delta a$ similar to those currently predicted for $a_\mu^{\rm SM}$. This scenario is indicated by the purple band in Figure~\ref{fig:reverseArgument}.

\subsection{The electron $g$-2}\label{sec:a_e}

Increasing $\sigma_{\rm had} (s)$ to account for $\Delta a_\mu$ also affects other observables. Here, its implications are considered for the electron magnetic moment anomaly, $a_e$, as well as the rescaled ratio of the leading hadronic vacuum polarization contributions to the electron and muon anomalies, $R_{e/\mu}$.

The electron anomaly currently exhibits a small but interesting discrepancy between SM theory and experiment. However, because hadronic loop effects scale as the squared lepton mass $m_\ell^2$, $a_e$ is less sensitive to them than $a_\mu$ (where they dominate the theory uncertainty). Instead, in the case of $a_e$, QED uncertainties induced by the error in the fine structure constant $\alpha$, along with the experimental value of $a_e$, dominate the uncertainties. Currently, the two most precise measurements of $\alpha$ arise from separate rubidium (Rb)~\cite{Bouchendira:2010es} and cesium (Cs) atomic interferometry~\cite{Parker:2018vye} experiments. These experiments are used to determine the electron mass very precisely via the ratios $h/M_{\rm Rb}$ and $h/M_{\rm Cs}$, respectively. Used in conjunction with the exceptionally well measured Rydberg constant, they provide the best determination of $\alpha$. Employing these two values and the recent re-evaluations of the hadronic vacuum polarization contributions to $a_e$~\cite{Keshavarzi:2019abf}, the SM predictions for the electron $g$-2 are~\cite{Keshavarzi:2019abf,Aoyama:2017uqe,Laporta:2017okg,Aoyama:2019ryr}
\begin{align}
a_{e}^{\rm SM}(\alpha_{\rm Rb}) & = (1159652182.042 \pm 0.72) \times 10^{-12} \ , \nonumber
\\
\
a_{e}^{\rm SM}(\alpha_{\rm Cs}) & = (1159652181.620 \pm 0.23) \times 10^{-12} \ .
\end{align}
Compared with the precise measurement of $a_e^{\rm exp} = (1159652180.73 \pm 0.28) \times 10^{-12}$~\cite{Hanneke:2008tm}, these two cases result in the following deviations between theory and experiment,
\begin{align}
\Delta a_{e}(\alpha_{\rm Rb}) & = (-1.31 \pm 0.77) \times 10^{-12} \ (1.7\sigma) \, , \nonumber
\\
\
\Delta a_{e}(\alpha_{\rm Cs}) & = (-0.89 \pm 0.36) \times 10^{-12} \ (2.5\sigma) \ .
\end{align} 
Currently, the uncertainties in these discrepancies are dominated by the errors of the experimental measurements of both $a_e$ and $\alpha$, rather than QED and hadronic effects.

Of interest is the current sign difference between $\Delta a_{e}$ and $\Delta a_{\mu}$ in conjunction with the relatively large magnitude of $\Delta a_{e}$. This may indicate the presence of new-physics contributions that may not scale naively with the square of lepton masses, as discussed in~\cite{Giudice:2012ms}. In that reference, examples of new-physics theories were presented in which the naive scaling $(m_e/m_{\mu})^2$ is violated inducing larger effects in the electron $g$-2. In such models, the value of $a_e$ was shown to be correlated with specific predictions for processes with violation of lepton number or lepton universality, and with the EDM of the electron. More recently, a single scalar solution to both electron and muon anomalies was shown to be possible if the two-loop electron Barr-Zee diagrams dominate the scalar one-loop electron anomaly effect, and the scalar couplings to the electron and two photons are relatively large~\cite{Davoudiasl:2018fbb} (for a detailed discussion of the contributions of spin-0 axion-like particles to lepton dipole moments, see~\cite{Marciano:2016yhf}). Combined explanations of the muon and electron $g$-2, and implications for a large muon EDM, were explored in~\cite{Crivellin:2018qmi}.

\begin{figure}[!t] 
\centering
\subfloat[The hadronic VP contributions, $a_{e}^{\rm had,\,VP} $. The solid-red line displays the curve obtained for point-like increases, with the uncertainty given by the orange band. The dashed-crimson, dashed-maroon and dashed-cyan lines represent the binned shifts for $\delta = 100,\,210,\,400$ MeV, respectively. The KNT19 result is given by the light-green band~\cite{Keshavarzi:2019abf}.]{%
\includegraphics[width= 0.47\textwidth]{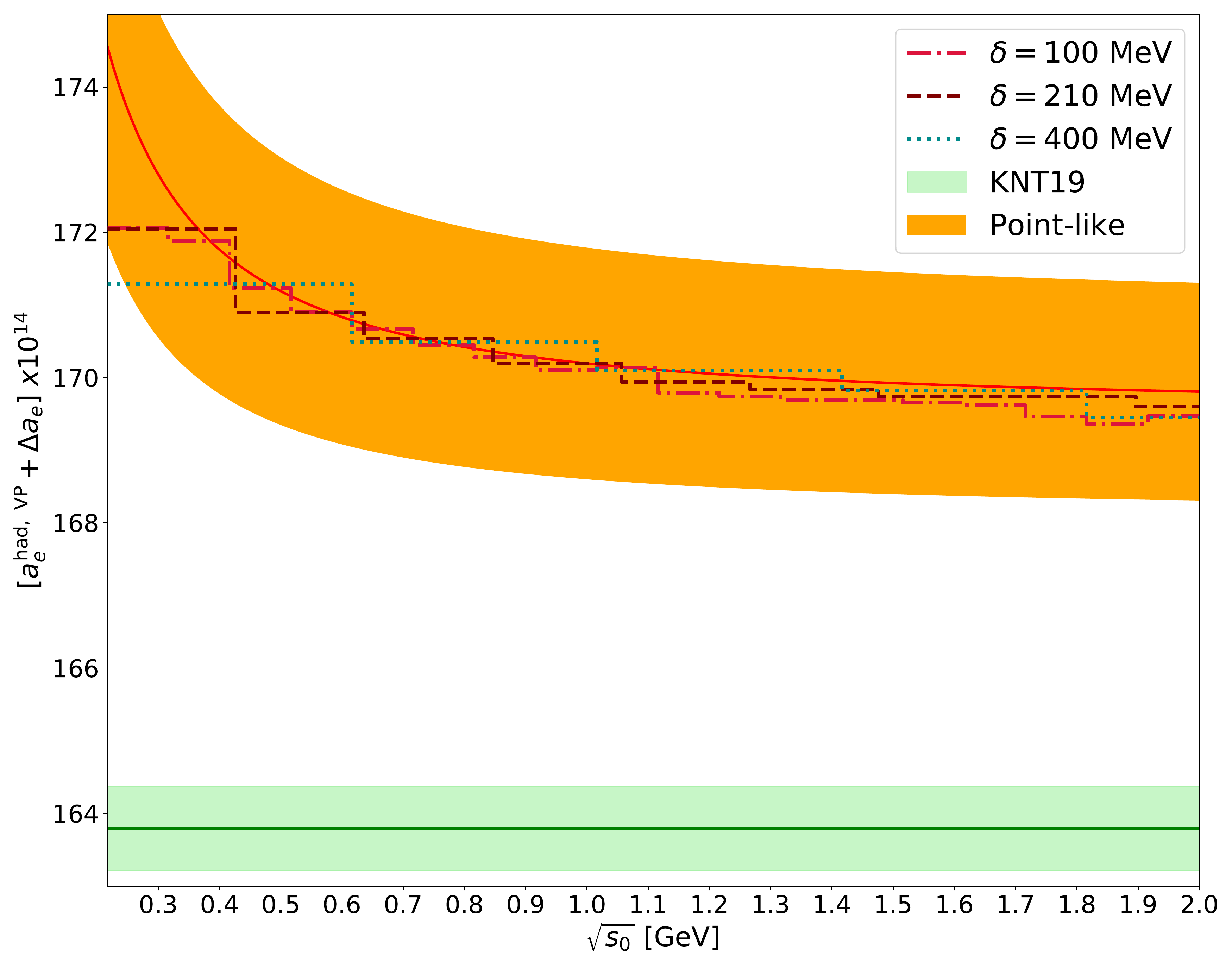}\label{fig:aeHVPS}}\hfill
\subfloat[The SM prediction, $a_e^{\rm SM}$. The solid-orange/solid-gray line gives the central values of $a_{e}^{\rm SM}(\alpha_{\rm Rb})$/$a_{e}^{\rm SM}(\alpha_{\rm Cs})$~\cite{Keshavarzi:2019abf}. The dashed-red/dashed-black line and light-yellow/dark-gray uncertainty band displays the curve obtained for point-like increases of $a_{e}^{\rm SM}(\alpha_{\rm Rb})$/$a_{e}^{\rm SM}(\alpha_{\rm Cs})$ The experimental measurement of $a_e$~\cite{Hanneke:2008tm} is given by the light-blue band.] {%
\includegraphics[width= 0.47\textwidth]{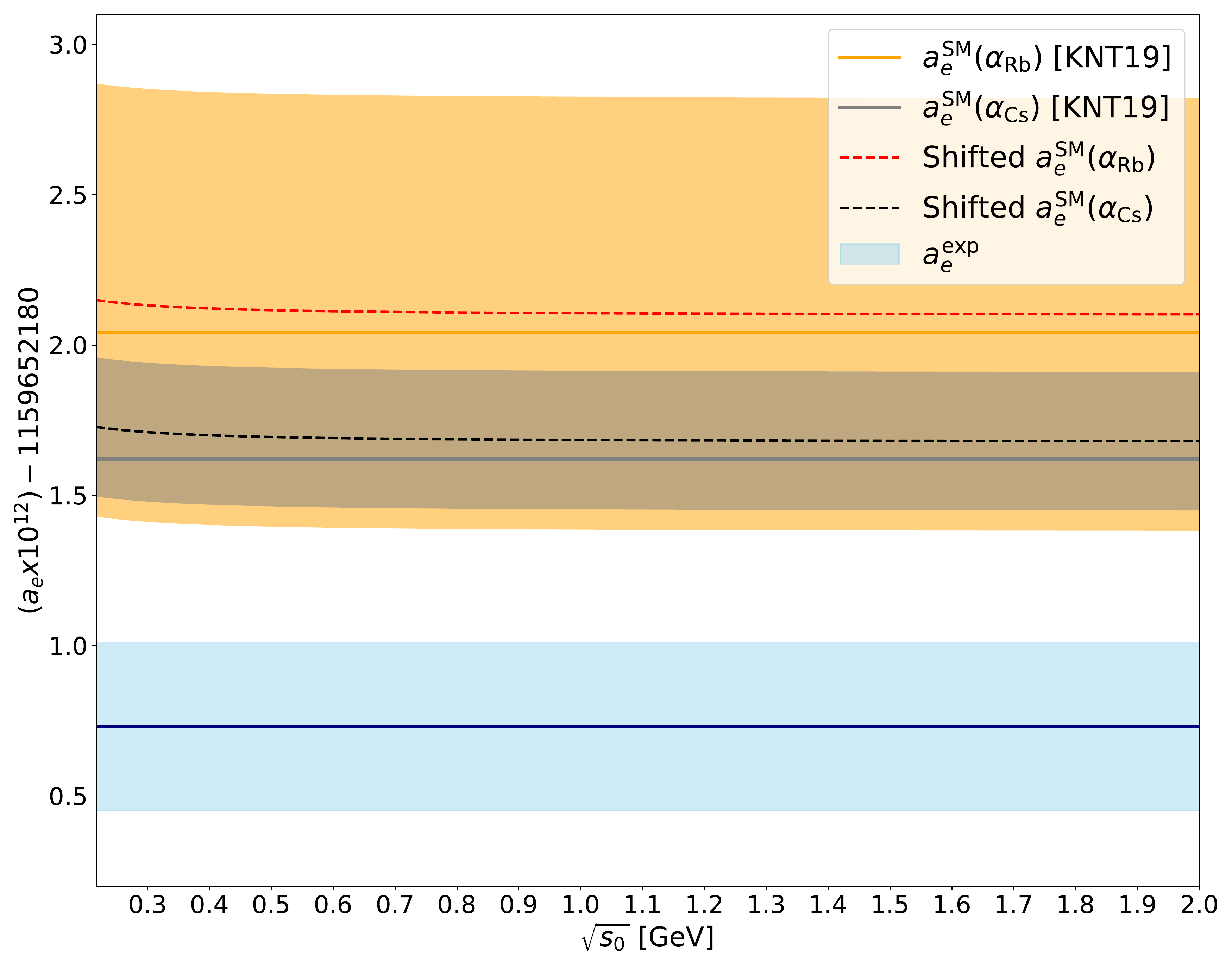}\label{fig:aeSM}}\hfill
\caption{\small Energy-dependent increases to $a_{e}$ observed when adjusting $\sigma_{\rm had}(s)$ to account for $\Delta a_{\mu}$.} \label{fig:a_e}
\end{figure}
The hadronic vacuum polarization contributions to $a_e$ are determined in an identical fashion to $a_{\mu}^{\rm had,\,VP}$, as described in Section~\ref{sec:dAlphaShiftMethod}, but with modified kernel functions that depend on the electron mass instead of the muon mass. This causes $a_{e}^{\rm had,\,VP} $ to be dominated by the contributions from lower energies, even more so than for the muon. Therefore, the shifts applied to $\sigma_{\rm had} (s)$ in Section~\ref{sec:Results} to account for $\Delta a_\mu$ induce larger relative increases to $a_{e}^{\rm had,\,VP} $ at lower energies than for $a_{\mu}^{\rm had,\,VP}$. However, as $a_e^{\rm SM}$ is less sensitive to the hadronic vacuum polarization sector than $a_\mu^{\rm SM}$, the influence of these on the comparison with $a_e^{\rm exp}$ is correspondingly weaker than in the case of the muon $g$-2. 

The impact of the shifts of $\sigma_{\rm had} (s)$ on the electron $g$-2, for both $a_{e}^{\rm had,\,VP} $ and $a_e^{\rm SM}$, is shown in Figure~\ref{fig:a_e}. For $a_{e}^{\rm had,\,VP} $, the expected emphasis at low energies is evident. Figure~\ref{fig:aeSM} displays the weakened effect on $a_e^{\rm SM}$, with only a slight increase visible close to the hadronic production threshold. When comparing the shifted results with their SM predictions, the overall significance of the observed increases is small. However, increasing $\sigma_{\rm had} (s)$ to account for $\Delta a_\mu$ always (slightly) increases the deviation $\Delta a_e$, invoking additional tension. For the $\alpha_{\rm Rb}$ determination, $\Delta a_{e}(\alpha_{\rm Rb})$ increases to $1.8\sigma$ at the $\sqrt{s_0}= m_{\pi^0}$ threshold. In the case of $\alpha_{\rm Cs}$, the deviation $\Delta a_{e}(\alpha_{\rm Cs})$ increases to $2.8\sigma$.\footnote{Strikingly, it should be noted that reversing this argument by decreasing $\sigma_{\rm had} (s)$ to fix $\Delta a_e$ results in a factor $\sim 12$ increase to $\Delta a_\mu$.}

The changes in both the muon and electron $g$-2 have a cumulative effect on the ratio of the leading hadronic vacuum polarization contributions to their anomalies:
\beq
R_{e/\mu} = \left(m_\mu/m_e \right)^2 \left(a_{e}^{\rm had,\,LO\,VP}/a_{\mu}^{\rm had,\,LO\,VP} \right)\, .
\eeq
The $\left(m_\mu/m_e \right)^2$ factor is introduced to cancel the leading $m_e/m_\mu$ dependence of \break  $a_{e}^{\rm had,\,LO\,VP}/a_{\mu}^{\rm had,\,LO\,VP}$, which is roughly of order $\left(\left(\alpha/\pi\right) \left(m_e/m_{\rho}\right)\right)^2/\left(\left(\alpha/\pi\right) \left(m_\mu/m_{\rho}\right)\right)^2$, where $m_\rho$ is the $\rho$ meson mass. This quantity was recently determined directly via lattice QCD to be $R_{e/\mu} = 1.1478 (70)$~\cite{Giusti:2020efo}. Taking the ratio of the KNT19 values from equation~\eqref{eq:KNT19HVP} and $a_{e}^{\rm had,\,LO\,VP}[{\rm KNT19}] = (186.08 \pm 0.66) \times 10^{-14}$~\cite{Keshavarzi:2019abf}, together with the CODATA value of the mass ratio of $m_\mu/m_e = 206.7682831 (47)$~\cite{Mohr:2015ccw}, results in $R_{e/\mu}[{\rm KNT19}] = 1.148343 (62)$ for 100\% correlated errors.

\begin{figure}[!t] 
\centering
\includegraphics[width=0.6\textwidth]{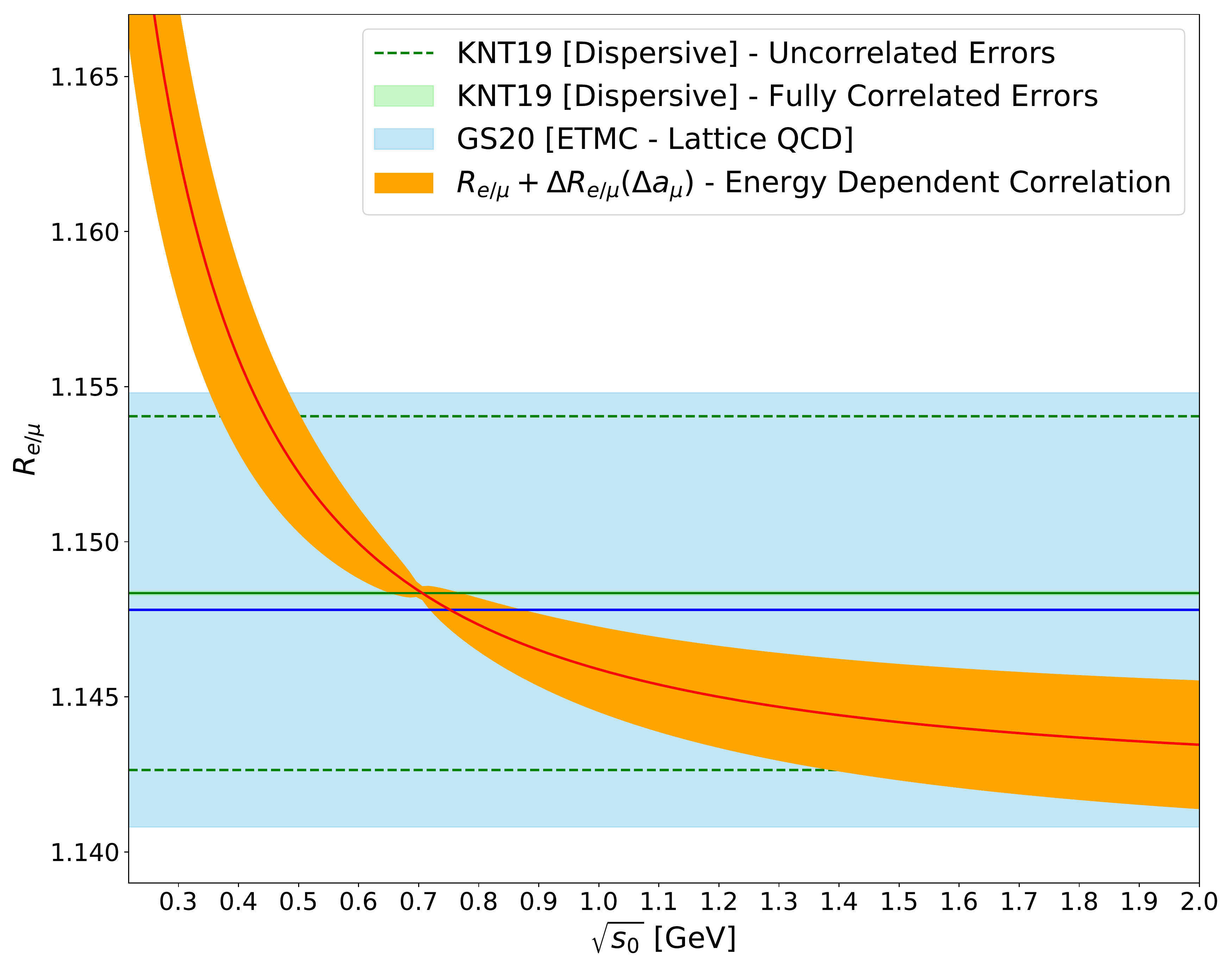}
\caption{\small $R_{e/\mu}$ obtained when adjusting $\sigma_{\rm had}(s)$ to account for $\Delta a_{\mu}$. The solid-red line displays the curve obtained for point-like increases, with the uncertainty given by the orange band. The result determined from the KNT19 values result is given by the light-green band (for 100\% correlated errors from $e$ and $\mu$) and the dashed-green lines (for uncorrelated errors from $e$ and $\mu$). The light-blue band shows the recent lattice QCD evaluation of~\cite{Giusti:2020efo}.} \label{fig:Remu}
\end{figure}
It follows that adjusting the cross section to account for $\Delta a_\mu$ results in the orange band shown in Figure~\ref{fig:Remu}, where the energy dependent correlation of the uncertainties on $a_{e}^{\rm had,\,LO\,VP}$ and $a_{\mu}^{\rm had,\,LO\,VP}$ is taken into account for the different values of $\sqrt{s_0}$. The resulting values for $R_{e/\mu}$, which are large for energies close to threshold and decrease for increasing $\sqrt{s_0}$, are compared with the result from~\cite{Giusti:2020efo} and the KNT19 value, which are in good agreement. Although the LQCD precision is not yet sufficient to make firm conclusions, it is interesting to note that the increased value of $R_{e/\mu}$ for shifts in the cross section at lower energies could provide an opportunity to place additional bounds on accounting for $\Delta a_\mu$ in $a_{\mu}^{\rm had,\,LO\,VP}$, should the lattice precision improve in the future. This is particularly relevant given the findings of Section~\ref{sec:Results}, where the results from the EW fit exclude increases to the cross section for higher energies.

\subsection{The weak mixing angle at low energies}\label{sec:weakangle}

Direct measurements and theoretical studies of the running weak mixing angle $\theta_W$ at low space-like $q^2=-Q^2<0$ also provide insight regarding hadronic vacuum polarization effects. They connect $\theta_W$ measured at low $Q^2$ in parity-violating polarized electron scattering and atomic physics asymmetry experiments~\cite{PDG2018}, with the more precisely $Z$-pole determined value at high $q^2=M_Z^2$. Hadronic vacuum polarization effects contribute to the evolution via $\gamma-Z$ mixing in much the same way as they influence the running of $\alpha(q^2)$ from $0$ to $M_Z^2$, but roughly with opposite sign. In both cases, the non-perturbative part of the running can be connected by a dispersion relation to $e^+e^- \rightarrow \text{hadrons}$ cross section data or calculated via lattice QCD. Of course, an increase of $\sigma_{\rm had}(s)$ to bridge the muon $g$-2 discrepancy also modifies the running of $\theta_W$. 

Although the running of the weak mixing angle can be described in terms of a physical, on-shell defined angle, it is simpler to make use of an ${\overline {\rm MS}}$ defined angle which has proved useful in discussing grand unified theories~\cite{Marciano:1980be}. In the ${\overline {\rm MS}}$ scheme, the running $\sin^2 \! \hat{\theta}_W(\mu)$ is the ratio of the QED coupling and the ${\rm {SU(2)}}_L$ SM gauge coupling, $\sin^2 \! \hat{\theta}_W(\mu)= \hat{\alpha}(\mu)/\hat{\alpha}_2(\mu)$, where $\mu$ is an arbitrary sliding mass scale~\cite{Marciano:1980be}. For $\mu = M_Z$, $\sin^2 \! \hat{\theta}_W(M_Z)$ can be obtained from the precisely measured effective weak mixing angle $\sin^2 \! \theta^{\rm lep}_{\rm eff}$ by the well-known relation~\cite{Gambino:1993dd,PDG2018}
\beq\label{eq:sin2thetaMZ}
\sin^2 \! \hat{\theta}_W(M_Z) =\sin^2 \! \theta^{\rm lep}_{\rm eff} - 0.00032.
\eeq
The quantity of direct interest for very low-energy experiments is $\sin^2 \! \hat{\theta}_W(0)$~\cite{Marciano:1980be, Marciano:1982mm, Jegerlehner:1985gq, Czarnecki:1995fw, Czarnecki:1998xc, Czarnecki:2000ic, Ferroglia:2003wa, Erler:2004in, Kumar:2013yoa}. It is obtained from $\sin^2 \! \hat{\theta}_W(M_Z)$ using the calculated quantum corrections induced by $\gamma-Z$ mixing and other radiative corrections~\cite{Czarnecki:1995fw, Ferroglia:2003wa, Erler:2004in}
\beq\label{eq:sin2theta0Erler}
\sin^2 \! \hat{\theta}_W(0) = \hat{\kappa}(0)\sin^2 \! \hat{\theta}_W(M_Z) \,,
\eeq
where $\hat{\kappa}(0) = 1.03196(6)$~\cite{Erler:2017knj}. That expression can be re-arranged to~\cite{Erler:2017knj}
\beq\label{eq:sin2theta0}
\sin^2 \! \hat{\theta}_W(0) \, = \, k\sin^2 \! \hat{\theta}_W(M_Z) \, + \, k'\sin^2 \! \hat{\theta}_W(M_Z)\int^{4 \, {\rm GeV}^2}_{s_{th}} \! {\rm d}s\,g(s)\,\sigma_{\rm had} (s) \,,
\eeq
where $k = 1.02527(4)$, $k' = 1.14$ and the integral in the last term corresponds exactly to equation~\eqref{eq:b}, except with an upper integration limit of $s = 4 {\rm GeV}^2$. The integral isolates the non-perturbative hadronic corrections, allowing for updates as well as for evaluations where $\sigma_{\rm had}(s)$ is increased to accommodate $\Delta a_{\mu}$. From the methodology described in Section~\ref{sec:dAlphaShiftMethod}, point-like increases to $\sigma_{\rm had}(s)$ alter equation~\ref{eq:sin2theta0} as:\footnote{Note that expressions corresponding to those given in Section~\ref{sec:dAlphaShiftMethod} are also derived for binned shifts and for the uncertainties on the point-like shifts.}
\beq\label{eq:sin2theta0'}
\sin^2 \! \hat{\theta}_W(0) \, \rightarrow \, \sin^2 \! \hat{\theta}_W(0) + k'\sin^2 \! \hat{\theta}_W(M_Z)\Delta a_\mu\frac{g(s_0)}{f(s_0)} \,.
\eeq

\begin{figure}[!t] 
\centering
\subfloat[Scenario (1).]{%
\includegraphics[width= 0.47\textwidth]{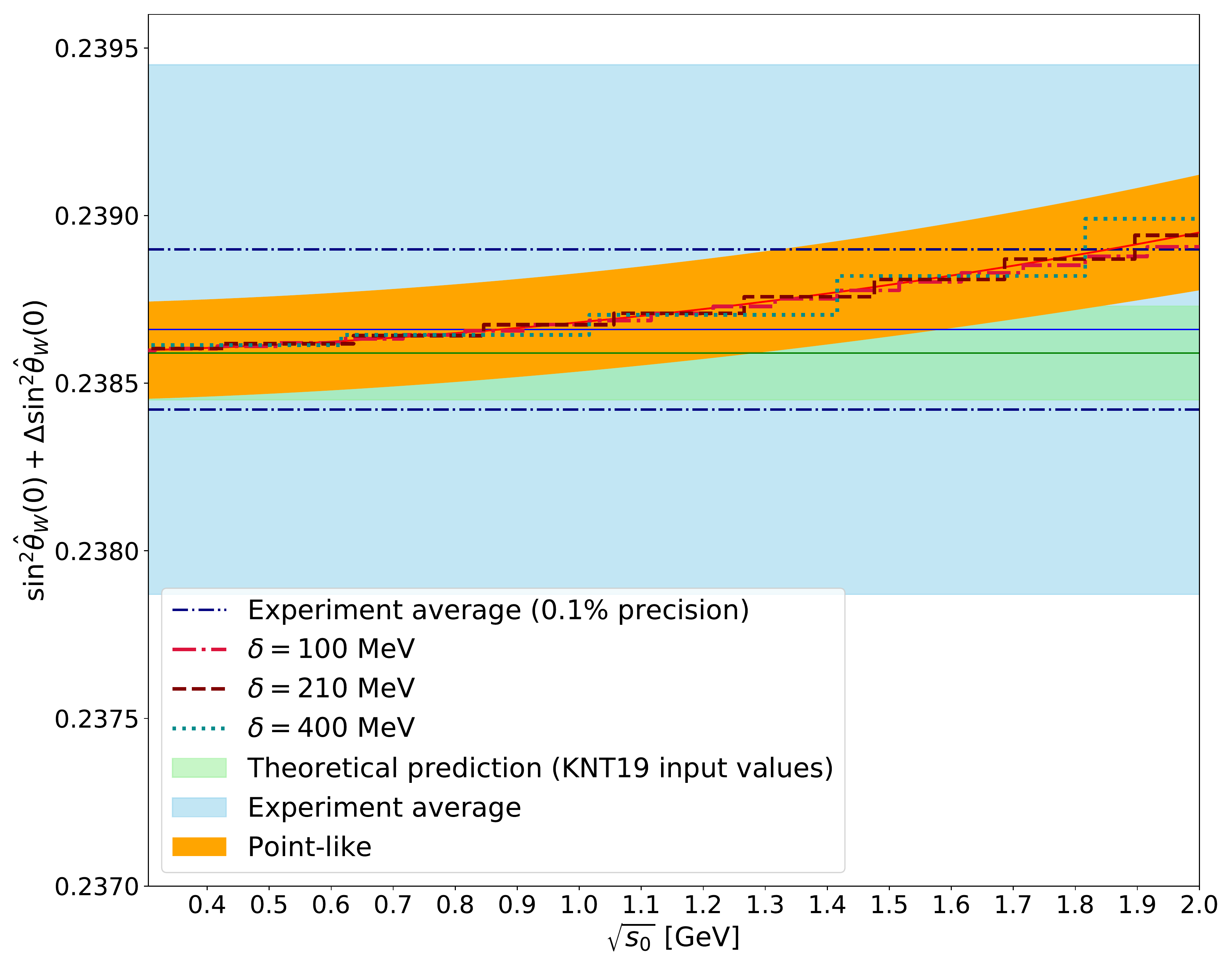}\label{fig:s20Shift}}\hfill
\subfloat[Scenario (2).] {%
\includegraphics[width= 0.47\textwidth]{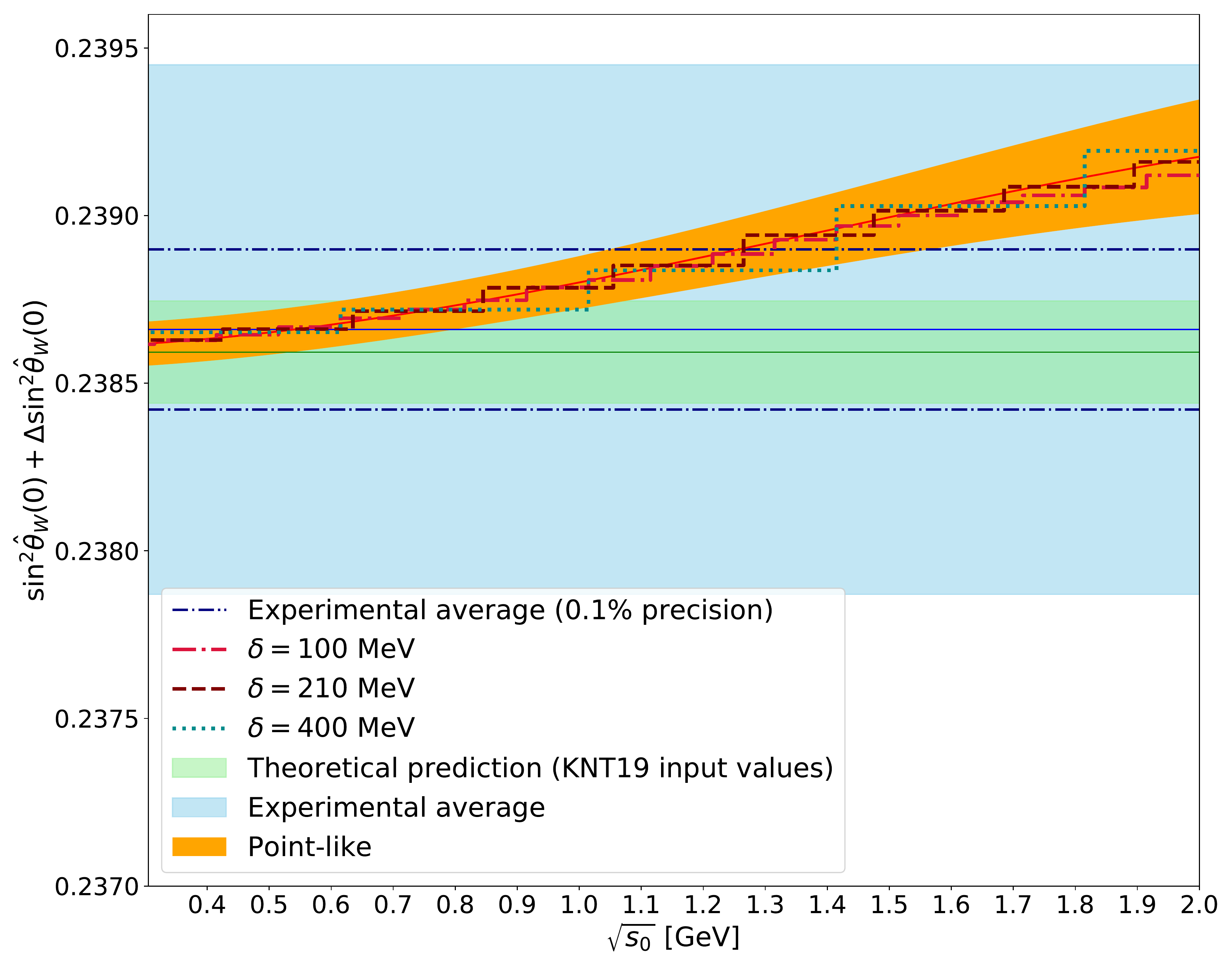}\label{fig:s20EWShift}}\hfill
\caption{\small Predictions for $\sin^2 \! \hat{\theta}_W(0)$ obtained adjusting $\sigma_{\rm had}(s)$ to account for $\Delta a_{\mu}$. Figure~\ref{fig:s20Shift} shows scenario (1), where the experimental measurement for $\sin^2 \! \theta^{\rm lep}_{\rm eff} $ is used as input into equation~\eqref{eq:sin2thetaMZ}. Figure~\ref{fig:s20EWShift} shows scenario (2), where the varied values of $\sin^2 \! \theta^{\rm lep}_{\rm eff}$ predicted from the EW fit are used as input into equation~\eqref{eq:sin2thetaMZ}. The solid-red line displays the curve obtained for point-like increases, with the $\pm1\sigma$ uncertainty given by the orange band. The dashed-crimson, dashed-maroon and dashed-cyan lines represent the binned shifts for $\delta = 100,\,210,\,400$ MeV, respectively. The theoretical prediction corresponding to KNT19 input values is given by the light-green band~\cite{Keshavarzi:2019abf}. The experimental average of $\sin^2 \! \hat{\theta}_W(0)$ obtained in Table~\ref{tab:sin20} is given by the light-blue band. The navy-blue bands depict the expected future uncertainty on the measured value~\cite{Becker:2018ggl,Benesch:2014bas}.} \label{fig:s20}
\end{figure}

The impact of the results of Section~\ref{sec:Results} on $\sin^2 \! \hat{\theta}_W(0)$ can be explored in two scenarios: (1) using $\sin^2 \! \theta^{\rm lep}_{\rm eff}$ as measured by experiment as a constant value in equation~(\ref{eq:sin2thetaMZ}), or (2) not using the experimental average for $\sin^2 \! \theta^{\rm lep}_{\rm eff}$ and employing instead the values of $\sin^2 \! \theta^{\rm lep}_{\rm eff}$ predicted by the global EW fit as shown in Figure~\ref{fig:EWvarST}. For scenario (1), using $\sin^2 \! \theta^{\rm lep}_{\rm eff} = 0.23151(14)$~\cite{Erler:2019hds}, equations~(\ref{eq:sin2thetaMZ}) and~(\ref{eq:sin2theta0Erler}) result in
\beq\label{eq:sin2theta0predicted}
\sin^2 \! \hat{\theta}_W(0) = 0.23858(15)\, . 
\eeq
From equation~\eqref{eq:sin2theta0'}, in this scenario, the results for $\sin^2 \! \hat{\theta}_W(0)$ are shown in Figure~\ref{fig:s20Shift}, where it can be seen that the increases to $\sigma_{\rm had}(s)$ to account for $\Delta a_\mu$ applied in Section~\ref{sec:Results} increase the value of equation~\eqref{eq:sin2theta0predicted} for all $\sqrt{s_0}$. The results for scenario (2) are shown in Figure~\ref{fig:s20EWShift}, which incorporates the values obtained for $\sin^2 \! \theta^{\rm lep}_{\rm eff}$ in Figure~\ref{fig:EWvarST}. In this case, for conservativeness, the orange band displays the bounds from the EW fit with the uncertainties from $\Delta\alpha_{\rm had}^{(5)}(M_{Z}^2)$ (corresponding to the integral in equation~\eqref{eq:sin2theta0}) and values of $\sin^2 \! \theta^{\rm lep}_{\rm eff}$ obtained from the EW fit taken to be 100\% correlated. It is evident that, in general, this scenario invokes a greater increase to $\sin^2 \! \hat{\theta}_W(0)$ than scenario (1), with smaller theoretical uncertainties for shifts at lower energies.

Experimental determinations of $\sin^2 \! \hat{\theta}_W(0)$ are not yet precise enough to either impose a significant constraint on accounting for $\Delta a_\mu$ in the hadronic polarization sector or provide a definitive test of the dispersion relation value for $\Delta\alpha_{\rm had}^{(5)}(M_{Z}^2)$. The current experimental status of $\sin^2 \! \hat{\theta}_W(0)$ is illustrated by considering the best three weak charge (vector weak neutral current couplings) $Q_W$ measurements, all taken at very low $Q^2$. These include the experimental results from atomic parity violation in Cs, $Q_W({\rm Cs})=-72.62(43)$, from parity-violating M{\o}ller scattering, $Q_W(e)=-0.0403(53)$, and from polarized $e^-p$ scattering, $Q_W(p)=0.0719(45)$~\cite{PDG2018}. After accounting for well-known SM quantum loop effects, results for the running ${\overline {\rm MS}}$ weak mixing angle (parametrized at 0 and at the $Z$-pole mass scale) are given in Table~\ref{tab:sin20}~\cite{PDG2018,Kumar:2013yoa}. The experimental average for $\sin^2 \! \hat{\theta}_W(0)$ determined from the weak charge inputs is 
\beq\label{eq:sin2theta0exp}
\sin^2 \! \hat{\theta}_W(0)^{\rm exp} = 0.23866(79)\, . 
\eeq
This value is in excellent agreement with the predicted one in equation~\eqref{eq:sin2theta0predicted}, although with a larger (0.33\%) uncertainty. It is shown by the light-blue bands in Figure~\ref{fig:s20}. The corresponding $Z$-pole value $\sin^2 \! \hat{\theta}_W(M_Z) = 0.23127(77)$ is in good agreement with Gfitter's global fit result of $\sin^2 \! \hat{\theta}_W(M_Z) =0.23120(4)$ obtained from equation~\eqref{eq:sin2thetaMZ}, in confirmation of SM radiative corrections.
\begin{table}[!t]
\vspace{-0.cm}
\centering
\scalebox{1.0}{
{\renewcommand{\arraystretch}{1.3}
\begin{tabular}{|c|c|c|}
\hline 
Weak charge measurement & $\sin^2 \! \hat{\theta}_W(0)$ & $\sin^2 \! \hat{\theta}_W(M_Z^2)$ \\
\hline 
$Q_W({\rm Cs})$ & $ 0.2356(20) $ & $ 0.2283(20) $ \\
$Q_W(e)$ & $ 0.2403(13) $ & $ 0.2329(13) $ \\
$Q_W(p) $ & $ 0.2384(11) $ & $ 0.2310(11) $ \\
\hdashline 
Experimental average & $ 0.23866(79) $ & $ 0.23127(77) $ \\
\hline 
\end{tabular} 
}
}\caption{Results and experimental averages of $\sin^2 \! \hat{\theta}_W(0)$ and $\sin^2 \! \hat{\theta}_W(M_Z)$ obtained from the values for the weak charge $Q_W$ measured in atomic physics and polarized electron parity-violating scattering reactions~\cite{PDG2018}.}
\label{tab:sin20}
\end{table}

Comparing SM theory and experiment provides the new-physics constraint $\delta \hat{\kappa}(0)^{\rm NP}= 0.0003(34)$~\cite{Czarnecki:1995fw}. Roughly speaking, this corresponds to $\Delta\alpha_{\rm had}^{(5)}(M_Z^2) = 0.0279(34)$, or approximately a $12$\% test of the hadronic contributions to the running. Although consistent with dispersion relation expectations, the error is currently too large to provide a definitive confirmation. It can also be translated into bounds on heavy $Z'$ bosons at roughly the 1~TeV level~\cite{Czarnecki:1995fw}. More generally, an anomalous electron anapole moment, $F_A(0)$, would cause a shift in $\sin^2 \! \hat{\theta}_W(0)$ of $-4\sqrt{2}\pi\alpha F_A(0)/G_F$ at a radius of about $10^{-17}$ cm (the $X$ parameter of dynamical symmetry breaking, which is constrained to $X <0.1$). In addition, there is the potential to probe other possibilities such as relatively light dark-$Z$ models~\cite{Davoudiasl:2012ag, Davoudiasl:2014kua, Davoudiasl:2015bua}. Planned experiments at MESA (P2) in Mainz~\cite{Becker:2018ggl} and JLab (M{\o}ller)~\cite{Benesch:2014bas} aim to reduce current errors by about a factor of 4 (depicted by the navy blue dashed lines in the plots of Figure~\ref{fig:s20}). At that level of precision, they will become competitive and complementary with $Z$-pole measurements for testing hadronic loop effects and probing for new physics. 

Alternative approaches are possible to test the relationships described here. The connection between $\sin^2 \! \theta_W(Q^2)$ and the QED coupling can be directly explored via lattice QCD calculations for space-like momenta. Consider the running QED and ${\rm SU(2)}_L$ gauge couplings. In running from $Q^2=0$ to $M_Z^2$, the former increases, due to hadronic loop effects, by $\Delta\alpha_{\rm had}^{(5)}(-M_Z^2)$, whereas the latter has hadronic loop corrections roughly twice that size. When their ratio is taken, i.e.\ the running $\sin^2 \! \theta_W$, one finds that the hadronic corrections are roughly $-\Delta\alpha_{\rm had}^{(5)}(-M_Z^2)$. Equality in magnitude holds for the light $u, d, s$ quarks in the SU(3) flavor limit for $\sin^2 \! \theta_W=1/4$. That feature has been observed in lattice QCD. For example, a recent 3-flavour LQCD calculation found for the hadronic part (ignoring disconnected diagrams) the related changes
$\Delta_{\rm had}\sin^2 \! \theta_W(-Q^2) \simeq -\Delta\alpha_{\rm had}(-Q^2)$ 
over the range $0<Q^2<5$~GeV$^2$, with $\Delta_{\rm had}\sin^2 \! \theta_W(-5 {\rm \, GeV}^2)=-0.006687(54)$ and $\Delta\alpha_{\rm had}(-5 {\rm \, GeV}^2)=0.006415(51)$~\cite{Ce:2019imp}. The roughly equal magnitudes and opposite signs are evident. The effect of $c$ and $b$ quarks can be calculated perturbatively. There are cancellations that enter such that $\Delta_{\rm had}\sin^2 \! \theta_W(-Q^2)$ and $\Delta\alpha_{\rm had}(-Q^2)$ also evolve with roughly equal magnitudes, but opposite sign. This feature is to some extent accidental, since it does not apply to lepton loops which have their contribution to the running of $\sin^2 \! \theta_W$ suppressed by $1-4\sin^2 \! \theta_W$. Of course, approximate equality in magnitude but opposite sign is interesting for rough discussions, but for precision comparisons a more complete calculation is needed. It must include additional EW radiative corrections, including $W$ boson loops. Employing the on-shell definition (correctly normalized to space-like $Q^2$), it should be possible to make the correspondence with the dispersive result $\Delta\alpha_{\rm had}^{(5)}(M_Z^2) = 0.02761(11)$ used throughout this work. Direct experimental constraints will also be possible by the proposed MUonE experiment at CERN~\cite{Calame:2015fva,Abbiendi:2016xup,MUonELOI,Banerjee:2020tdt}, where fixed-target $\mu$-$e$ scattering will facilitate a direct measurement of the hadronic vacuum polarization as a function of space-like $q^2$, providing a more direct test of $\alpha(M_Z^2)$.

\section{Conclusions}\label{Conclusions}

This analysis has updated and improved the study initiated in~\cite{Passera:2008jk}. Namely, it has examined the possibility that an underestimate of the hadronic vacuum polarization loop correction is responsible for the muon magnetic moment anomaly difference between SM theory and experiment. In particular, it has explored the feasibility that $\Delta a_\mu$ is due to hypothetical missed contributions in the total $e^+e^- \rightarrow \text{hadrons}$ cross section $\sigma_{\rm had}(s)$ that is used as input into dispersion relations to calculate $a_{\mu}^{\rm had,\,VP}$. The same cross section data are used as input into dispersion relations to calculate the hadronic contribution to the running QED coupling evaluated at the $Z$-pole, $\Delta\alpha_{\rm had}^{(5)}(M_{Z}^2)$, which is an important component of the global fits to the EW sector of the SM. By applying hypothetical changes to the cross section to account for $\Delta a_{\mu}$, the corresponding shifted values for $\Delta\alpha_{\rm had}^{(5)}(M_{Z}^2)$ have been input into the global EW fit to obtain accurate predictions for the $W$ boson mass $M_W$, the effective EW mixing angle $\sin^2 \! \theta^{\rm lep}_{\rm eff}$ and the Higgs boson mass $M_H$.

Employing improvements in $\sigma_{\rm had}(s)$ and the EW parameter fits (including the precisely measured Higgs boson mass) leads to improved constraints beyond those found in~\cite{Passera:2008jk}. For the $W$ mass, any increase in $\sigma_{\rm had}(s)$ to account for $\Delta a_{\mu}$ decreases the prediction of $M_W$ further away from its measured value. Accounting for $\Delta a_\mu$ in the hadronic vacuum polarization contributions is found here to be excluded for $\sqrt{s_0} \gtrsim 0.9$~GeV at the 95\%CL. For $\sin^2 \! \theta^{\rm lep}_{\rm eff}$, adjustments to $\sigma_{\rm had}(s)$ are consistent with the measured $\sin^2 \! \theta^{\rm lep}_{\rm eff}$ for all considered values of $\sqrt{s_0}$. In the case of the Higgs boson, its predicted mass is found to decrease, contrary to experiment, for larger values of $\Delta\alpha_{\rm had}^{(5)}(M_{Z}^2)$. As a result, shifts in the hadronic cross section needed to account for $\Delta a_{\mu}$ are found to be excluded for $\sqrt{s_0} \gtrsim 0.7$~GeV at the 95\% CL. 

Possible shifts of the hadronic cross section to bridge the muon $g$-2 discrepancy have been further examined at energies lower than $\sim 0.7$~GeV. For a chosen region ranging from threshold up to $\sim 0.7$~GeV, the multiplicative scale factor required to uniformly adjust $\sigma_{\rm had}(s)$ to account for $\Delta a_\mu$ has been found to be $\epsilon \approx +9\%$. This region is dominated by the $\pi^+\pi^-$ channel, particularly by the $\rho$-resonance, where missed contributions are unlikely given the large number of precise data sets that now exist for the $\rho$ in the $\pi^+\pi^-$ final state. Of deeper concern are the threshold contributions of the $\pi^+\pi^-$, $\pi^0\gamma$, $\pi^+\pi^-\pi^0$ and $\eta\gamma$ channels. These are estimated from ChPT, potentially indicating (without confirmatory experimental measurements of these contributions) that there is a possibility the theory predictions may be incomplete, consequently resulting in $\Delta a_\mu$. When considering the current $3.8\sigma$ tension between the KNT19 dispersive result for $a_{\mu}^{\rm had,\,VP}$~\cite{Keshavarzi:2019abf} and the recent lattice QCD evaluation of the BMW collaboration that indicates agreement with $a_{\mu}^{\rm exp}$~\cite{Borsanyi:2020mff}, this could suggest that the lattice calculation may have captured contributions below $\sim0.7$~GeV that are absent from the hadronic data (see e.g.~\cite{deRafael:2020uif}). A study comparing the results from BMW collaboration with the dispersive results is beyond the scope of this work and a detailed discussion can be found in~\cite{Aoyama:2020ynm,Lehner:2020crt}. However, it has been found here that the size of the missed contributions required to bridge the muon $g$-2 discrepancy (or the BMW-KNT19 discrepancy) would need to be improbably large given the robust status of the hadronic cross section measurements. 

Predicting $\Delta\alpha_{\rm had}^{(5)}(M_{Z}^2)$ from the EW fit results in a lower value than the estimate determined from the $\sigma_{\rm had}(s)$ data, resulting in a difference $\Delta[\Delta\alpha_{\rm had}^{(5)}(M_{Z}^2)]$ between the two predictions. In a study new to this work, adjustments have been made to $\sigma_{\rm had}(s)$ to account for $\Delta[\Delta\alpha_{\rm had}^{(5)}(M_{Z}^2)]$ and investigate the impact on $\Delta a_\mu$. Although the corresponding values for $\Delta a_\mu$ are not statistically significant due the large errors arising from the EW fit, all the obtained values prefer a larger $\Delta a_{\mu}$ difference than the current muon $g$-2 discrepancy found in~\cite{Keshavarzi:2019abf}.

The effect on $a_e$ and $\sin^2 \! \hat{\theta}_W(0)$ due to these increases to $\sigma_{\rm had} (s)$ has also been scrutinized. For $a_e$, due to the current sign difference observed between $\Delta a_{e}$ and $\Delta a_{\mu}$, bridging the muon $g$-2 discrepancy in the hadronic vacuum polarization sector results in additional tension. This analysis has also revealed the potential for future bounds to be set via the ratio of the electron and muon hadronic vacuum polarization contributions, $R_{e/\mu}$, which was recently determined from lattice QCD~\cite{Giusti:2020efo}. For the weak mixing angle at low energies, the adjustments to $\sigma_{\rm had} (s)$ increase the value of $\sin^2 \! \hat{\theta}_W(0)$ for all considered $\sqrt{s_0}$. Although the experimental precision of $\sin^2 \! \hat{\theta}_W(0)$ is insufficient to impose significant additional constraints at this time, it has been shown that competitive improvements in this regard have the potential to test new physics scenarios which could enter the running of the weak mixing angle at low energies.

The prospects for alternative confirmations of the dispersive estimates of $a_{\mu}^{\rm had,\,VP}$ rest with either additional lattice QCD calculations or direct experimental measurement, as proposed by the MUonE experiment~\cite{Calame:2015fva, Abbiendi:2016xup, MUonELOI, Banerjee:2020tdt}. This experiment has been recently proposed at CERN to provide a new direct determination of the leading hadronic contribution to the muon $g$-2 measuring $\Delta\alpha_{\rm had}^{(5)}(q^2)$ for space-like values of $q^2$ via muon-electron scattering. Its future results could therefore help in understanding the present intriguing dichotomy between the dispersive evaluations of $a_{\mu}^{\rm had,\,LO\,VP}$ and the recent lattice QCD result from the BMW collaboration~\cite{Borsanyi:2020mff}

The fate of the $\Delta a_\mu$ discrepancy should soon be decided by the new Muon $g$-2 experiment at Fermilab and the follow-up experiment at  J-PARC. If $a_{\mu}^{\rm exp}$ should come into agreement with the SM prediction based on $a_{\mu}^{\rm had,\,VP}$ calculated from dispersion relations, it will mark the end of an era that has strongly challenged theoretical creativity and computational innovation. Alternatively, confirmation of the discrepancy at the $\geq 5\sigma$ level will strengthen the new physics interpretation and invigorate the quest for its underlying origin.

\section*{Acknowledgments}

We would like to thank G.~Degrassi, F.~Feruglio, P.~Gambino, P.~Giardino, M.~Hoferichter, M.~Lancaster, and T.~Teubner for very useful discussions. Special thanks are extended to D.~Nomura and T.~Teubner for their collaboration with A.K.\ in producing the compilation of hadronic cross section data used here. The work of A.K.\ was supported by STFC under the consolidated grant ST/S000925/1. The work of W.J.M.\ was supported by the U.S.\ Department of Energy under Grant DE-SC0012704. The work of A.S. was supported in-part by the National Science Foundation under Grant PHY-1915219. 

Work supported by the Fermi National Accelerator Laboratory, managed and operated by Fermi Research Alliance, LLC under Contract No. DE-AC02-07CH11359 with the U.S. Department of Energy. The U.S. Government retains and the publisher, by accepting the article for publication, acknowledges that the U.S. Government retains a non-exclusive, paid-up, irrevocable, world-wide license to publish or reproduce the published form of this manuscript, or allow others to do so, for U.S. Government purposes.


\end{document}